\documentclass[twocolumn,epjc3]{svjour3}
\RequirePackage[T1]{fontenc}
\smartqed
\usepackage{epsfig}
\usepackage{slashed}
\usepackage{graphicx}
\usepackage{multirow}
\usepackage{color}
\usepackage{balance}
\usepackage{appendix}
\usepackage{caption}
\usepackage{subcaption}
\usepackage{amsmath}
\usepackage{mathrsfs}
\usepackage{bm}
\usepackage{braket}
\usepackage{booktabs}
\usepackage{array}
\usepackage[mathscr]{euscript} 
\usepackage{dcolumn}      
\usepackage{slashed}       

\usepackage{mathptmx}
\RequirePackage[numbers,sort&compress]{natbib}
\RequirePackage[colorlinks,citecolor=blue,urlcolor=blue,linkcolor=blue]{hyperref}

\RequirePackage{graphicx}
\RequirePackage{amsmath}
\RequirePackage[cal=cm,scr=boondoxo,scrscaled=1.05]{mathalfa}  

\journalname{Eur. Phys. J. C}

\graphicspath{{figure-2/},{figure/}}
\begin{document}  	
	\title{Production of axion-like particles via vector boson fusion at future electron-positron colliders}
	
	\author{ Chong-Xing Yue \thanksref{e1,addrA}\and
		Hua-Ying Zhang \thanksref{e2,addrA} \and
		Han Wang \thanksref{e3,addrA}
}
	
	\thankstext{e1}{e-mail: cxyue@lnnu.edu.cn  }
	\thankstext{e2}{e-mail: huayingheplnnu@163.com \emph{\rm(corresponding ~author)} }
	\thankstext{e3}{e-mail: wangwanghan1106@163.com}
	
	\institute{Department of Physics, Liaoning Normal University, Dalian 116029, China\label{addrA}}
	

\date{Received: date / Revised version: date}
 \maketitle
\abstract{
One kind of particularly interesting pseudoscalar particles, called axion-like particles (ALPs), have rich physical phenomenology at high- and low-energy collider experiments. After discussing most of single production channels of ALP at electron-positron colliders, we investigate the possibility of detecting this kind of new particles through the W$^{+}$W$^{-}$ fusion process e$^{+}$e$^{-}$ $\rightarrow$  $\overline{\nu}_{e}$$\nu_{e}a$ $(\rightarrow \gamma \gamma)$ at the CLIC. The 3$\sigma$ and 5$\sigma$ bounds on the ALP parameter space at the three energy stages of the CLIC are obtained. We find that the bounds given by the CLIC are complementary to the existing experiments exclusion regions.
\PACS{
} 
} 

\newpage
\section{Introduction}\label{sec:intro}

\indent

After discovery of the Higgs boson in the CMS and ATLAS experiments~\cite{Aad:2012tfa,Chatrchyan:2012ufa}, experiments at the LHC have made great achievements and demonstrated that the standard model (SM) is a correct model to explain most of observed phenomena at the electroweak scale. However, deviations of experimental measurements from the SM predictions at or below the percent level are beyond the achievable precision at the LHC and is expected to be performed at the planned e$^{+}$e$^{-}$ colliders with high luminosity and energy.
It is well known that, compared to the LHC, the future e$^{+}$e$^{-}$ colliders, such as the  ILC~\cite{Djouadi:2007ik,Baer:2013cma,Moortgat-Picka:2015yla,Aihara:2019gcq,Bambade:2019fyw}, CLIC~\cite{Battaglia:2004mw,Roloff:2018dqu}, FCC-ee~\cite{Gomez-Ceballos:2013zzn,Abada:2019zxq}, and CEPC~\cite{CEPCStudyGroup:2018rmc,CEPCStudyGroup:2018ghi,An:2018dwb} have higher luminosity and more clean experimental environment, which can not only study the SM observables at unprecedented accuracy, but would be very useful to discover the evidence of new physics beyond the SM.

The discovery of the Higgs boson has stimulated interest in searching for additional (pseudo) scalar particles in high- and low-energy experiments. Axion-like particles (ALPs) are pseudoscalar bosons that naturally appear in many extensions of the SM as pseudo Nambu-Goldstone bosons arising in explicit global symmetry breaking~\cite{Peccei:1977hh,Weinberg:1977ma,Wilczek:1977pj,Kim:1979if,Shifman:1979if,Dine:1981rt,Zhitnitsky:1980tq}. The properties of ALPs, such as their masses and coupling strengths to the SM particles, are model-dependent and have been extensively investigated. High-energy collider experiments provide possibilities to explore ALPs with masses about $M_{a}$ $\geq$ 1 GeV~\cite{Bauer:2017ris,Bauer:2018uxu,Fortin:2021cog,dEnterria:2021ljz,Agrawal:2021dbo,Brivio:2017ije,Knapen:2016moh,Baldenegro:2019whq,Coelho:2020saz,Coelho:2020syp,Goncalves:2021pdc}. The constraints on a wide range of the ALP parameter space are obtained using the available data, such as the LEP, Tevatron and LHC data, the future prospects of discovering ALPs at running or future collider experiments are studied. For instance, searches for ALPs via photon fusion are investigated at the LHC and e$^{+}$e$^{-}$ colliders~\cite{Jaeckel:2012yz,Baldenegro:2018hng,Florez:2021zoo,Marciano:2016yhf,Inan:2020aal,Inan:2020kif,Zhang:2021sio,Buttazzo:2018qqp,Gavela:2019cmq}. The capabilities of these colliders to probe the ALP parameter space are discussed and the expected bounds on the ALP-photon coupling for a wide range of masses are obtained.
 
ALPs can also be produced via massive vector boson fusion (VBF) processes (i.e. W$^{+}$W$^{-}$-fusion and ZZ-fusion processes) at high-energy e$^{+}$e$^{-}$ colliders~\cite{Georgi:1986df}. Since the production cross sections of these processes depend logarithmically on the center-of-mass (c.m.) energy, and are generally smaller than those of other processes at low-energy e$^{+}$e$^{-}$ colliders, there are very few studies about ALP productions via VBF processes with V = W$^{\pm}$, Z. In this work, we will consider the possibilities of detecting ALPs via VBF processes at high-energy e$^{+}$e$^{-}$ colliders like the CLIC. We will focus on ALPs that only couple to the electroweak gauge bosons and perform a careful investigation of the ALP signals at high-energy e$^{+}$e$^{-}$ colliders from VBF processes. The relevant Feynman diagrams are shown in Fig.~\ref{feynman-4}. Our numerical results show that VBF processes can extend the e$^{+}$e$^{-}$ sensitivity to a region of the ALP parameter space that is not covered by other experiments.

The rest of this paper is organized as follows. After summarizing the effective description of ALP interactions with the electroweak gauge bosons, we calculate and compare the production cross sections of ALPs via e$^{+}$e$^{-}$ annihilation and VBF processes at e$^{+}$e$^{-}$ colliders in Sec.~\ref{theoretical}. Based on the details of the analysis of the ALP signals and the relevant SM backgrounds from W$^{+}$W$^{-}$ fusion process, sensitivity projections of the CLIC to the ALP parameter space are presented in Sec.~\ref{analyze}. We report our conclusions in Sec.~\ref{conclusions}.

\section{Single production of ALP at e$^{+}$e$^{-}$ colliders } \label{theoretical}

\indent

The ALP interactions with the SM fermions and gauge bosons arise through five-dimensional operators, and their masses can  be treated independently of their couplings~\cite{Georgi:1986df}, which can be described via an effective Lagrangian. The effective interactions of ALP with electroweak gauge bosons, which are related  to our calculation, are given by the following dimension-5 effective Lagrangian~\cite{Brivio:2017ije}:
\begin{eqnarray}\label{1L}
	\label{lagrangian}
	\begin{aligned}
		\mathcal{L}_{\text {eff }}^{D \leq 5}&= \frac{1}{2}\left(\partial_{\mu} a\right)\left(\partial^{\mu} a\right)-\frac{M_{a}^{2}}{2} a^{2} +  C_{W W} \frac{a}{f_{a}} W_{\mu \nu}^{A} \tilde{W}^{\mu \nu, A}\\
		&+  C_{B B} \frac{a}{f_{a}} B_{\mu \nu} \tilde{B}^{\mu \nu},
	\end{aligned}
\end{eqnarray}
where ${ X}_{\mu\nu}$ denotes the field strength tensor for $SU(2)_L$ or $U(1)_Y$, ${\tilde X}^{\mu\nu}=\frac{1}{2} \varepsilon^{\mu \nu \alpha \beta} X_{\alpha \beta}$ with $\varepsilon^{0123}=1$ and $X\in\{B, W\}$. The ALP field and mass are denoted by $a$ and $M_{a}$ , respectively. The interaction of the ALP with two photons is contributed by the last two terms of Eq.~\ref{1L}. The dimensionful couplings $g_{a\gamma\gamma}$, $g_{aWW}$, $g_{a\gamma Z}$ and $g_{aZZ}$, control the coupling strength of the interactions between ALP and electroweak gauge bosons ($W^{\pm}$, Z, $\gamma$), which are given by
\begin{eqnarray}
\begin{aligned}
	\centering
  \quad \quad &g_{a\gamma\gamma}=\frac{4}{f_{a}}(c_{\theta_{W}}^{2} C_{BB}  + s_{\theta_{W}}^{2} C_{WW}) ,\\
 & g_{aWW}=\frac{4}{f_{a}}C_{WW}, \quad ~~~~~~~~~~~~ \\
 &g_{aZZ}=\frac{4}{f_{a}}(s_{\theta_{W}}^{2}C_{BB}+c_{\theta_{W}}^{2}C_{WW}), \\
 & g_{a\gamma Z}=\frac{4}{f_{a}}s_{2\theta_{W}}(C_{WW}-C_{BB}),
 \end{aligned}
\end{eqnarray}
where $s_{\theta_{W}}=\sin \theta_{W}$, $c_{\theta_{W}} = \cos \theta_{W}$ with $\theta_{W}$ being the weak mixing angle.
To simplify our analysis, we will assume C$_{WW}$ = C$_{BB}$ in this work and there are $g_{a\gamma Z}$ = 0, $g_{a\gamma \gamma}$ = $g_{aWW}$ = $g_{aZZ}$\footnote{In this case, Ref.~\cite{Wang:2021uyb} has investigated searching for ALPs via ALP-strahlung process at the LHC.}. In this case, ALPs can be singly produced via e$^{+}$e$^{-}$ annihilation and VBF processes at e$^{+}$e$^{-}$ colliders.
We use \textsf{FeynRules}~\cite{Alloul:2013bka} to generate the UFO model file corresponding to the effective Lagrangian Eq.~\ref{1L}. The cross sections of e$^{+}$e$^{-}$ annihilation and VBF processes at e$^{+}$e$^{-}$ colliders are calculated, which are shown in Fig.~\ref{sigmavsga}.
One can see from Fig.~\ref{sigmavsga} that the ALP-strahlung processes ($\gamma a$ and $Za$) are the main production channels. However, their cross sections become independent of $s$ in the high-energy limit $M^{2}_{a}\ll s$ and thus saturate quite soon and flatten out. The cross sections of VBF processes depend only logarithmically on $s$. Therefore, the cross section of W$^{+}$W$^{-}$ fusion process grows with $s$ increasing and can reach sizable values at high energies. In principle, the cross section of the $ZZ$ fusion process  follows the same trend as that of the W$^{+}$W$^{-}$ fusion process, while its value is about an order of magnitude smaller than that of the W$^{+}$W$^{-}$ fusion process, due to the fact that the neutral current couplings are smaller than the charged current couplings. The $\gamma\gamma$ fusion process has been extensively discussed in Ref. \cite{Zhang:2021sio}. The cross section of the ZZ fusion process e$^{+}$e$^{-}$ $\rightarrow$ e$^{+}$e$^{-}$$a$ is the smallest. In Ref.~\cite{Azzurri:2021nmy}, the similar conclusion is given by exploring the mass and cross section of Higgs with ultimate precision at the FCC-ee.

ALPs have been investigated through the ALP-strahlung processes e$^{+}$e$^{-}$ $\rightarrow$ $V$$a$  $(a\rightarrow \gamma \gamma)$, where $V$ = $\gamma, Z$ \cite{Bauer:2018uxu,Jaeckel:2015jla,Mimasu:2014nea}.
In Ref. \cite{Bauer:2018uxu}, the ALP-strahlung processes have also been studied at the CLIC, but the author assumes $C_{WW}$ = 0, considering $\gamma Za$ coupling, which is opposite to our work.  There are previous works about the ALP single production induced by two-photon fusion in future electron-positron colliders \cite{Inan:2020aal,Inan:2020kif,Zhang:2021sio}.
Thus, in this paper, we only consider the production of ALP via W$^{+}$W$^{-}$ fusion process e$^{+}$e$^{-}$ $\rightarrow$  $\overline{\nu}$$_{e}$$\nu$$_{e}$$a$ with $a$ subsequently decaying to photon pairs (see Fig.~\ref{signal-alp-decay}).

\section{Search for ALP at the CLIC} \label{analyze}

\indent

As we know, the high-energy e$^{+}$e$^{-}$ collider can provide clean experimental environment to explore new physics beyond the Standard Model (BSM).
The future e$^{+}$e$^{-}$ colliders are contemplated to study the properties of the SM observables with an unprecedented precision, and the CLIC is one of them.
The CLIC is planned to be built and operated in three energy stages. The first stage
 is planned to accurately measure the properties of the Higgs boson at $\sqrt{s}$ = 380 GeV with $\mathcal{L} $ = 0.5 ab$^{-1}$.
The second and ultimate stages running at $\sqrt{s}$ = 1500 GeV with $\mathcal{L} $ = 1.5 ab$^{-1}$ and $\sqrt{s}$ = 3000 GeV with $\mathcal{L} $ = 3 ab$^{-1}$ are not only able to measure the top Yukawa coupling, the Higgs trilinear self-coupling and the Higgs rare decays, but also can  give indirect sensitivity to many new physics scenarios through precision measurements~\cite{CLIC:2016zwp}. They will be able to discover new particles through direct detection. We will do calculations and analysis of the signal and background for the process $e^{+}e^{-} \rightarrow \overline{\nu}_{e}\nu_{e}a (\to\gamma\gamma$) at the CLIC with three energy stages.

In this section, we perform the Monte Carlo (MC) simulation to analyze the possibility of exploring ALPs at the CLIC through the W$^{+}$W$^{-}$ fusion process e$^{+}$e$^{-}$ $\rightarrow$ $\overline{\nu}_{e}$$\nu_{e}a$ $(\rightarrow \gamma \gamma)$, where the gauge bosons W$^{\pm}$ are emitted from the incoming electron/positron beams. Its signal is characterised by two photons, missing transverse energy from the escaping undetected two neutrinos ($\overline{\nu}$$_{e}$$\nu$$_{e}$). In our numerical calculations, we take into account the process e$^{+}$e$^{-}$ $\rightarrow$ $\overline{\nu}$$\nu$$\gamma\gamma$ as the dominant SM background, which is mainly induced through electroweak interaction. For the signal, the final two photons mainly come from the decay of pseudo-scalar ALPs (spin-0), while there is no particle with spin-0 in the background that decays to two photons at tree level. So the interference is very small.
We have estimated the cross section of pure ALP signal to be 17.25 fb and the cross section of the interference to be -0.1549 fb at $\sqrt{s}=1500$ GeV, $M_{a}=10$ GeV and $g_{a\gamma\gamma}=10^{-3}$ GeV$^{-1}$ without any cut applied. Apparently the cross section of pure ALP signal is two orders of magnitude larger than that of the interference.
We have estimated the cross sections of pure ALP signal and interference for multiple ALP benchmark points, and the conclusions are similar. For simplicity, we have neglected the influence of the interference between ALP production and SM background in $e^{+}e^{-}\to\overline{\nu}\nu\gamma \gamma$.

\begin{table*} \scriptsize 
	\centering{
		\caption{The improved cuts on the signal and background events. \label{cuts}}
\begin{tabular}{c|c c c }
\hline
	Cuts     &$\sqrt{s} = $ 380 GeV & $\sqrt{s} =$ 1500 GeV& $\sqrt{s} =$ 3000 GeV\\
\hline
 Cut-1:  Angle between the ALP and the beam axis & 0.6 $\textless ~\theta(\gamma \gamma)~ \textless$~2.6 & 0.7~$\textless~ \theta(\gamma \gamma) ~\textless$~2.5 &0.5~$\textless~\theta(\gamma \gamma) ~\textless$~2.6 \\
 Cut-2:  ~~~~~~~ ~ The missing transverse energy~~ ~~~~~~~~~~  &$\slashed{E}_{T} ~ \textgreater~70 ~\mathrm{GeV}$ &$\slashed{E}_{T}~  \textgreater~150 ~\mathrm{GeV}$   &$\slashed{E}_{T} ~ \textgreater~200 ~\mathrm{GeV}$ \\	

 Cut-3:  ~~~~~~~~~~~~~~~ The transverse energy~~~~~~~~~~~~~~~~~~~   & $E_{T}~\textgreater~85 ~\mathrm{GeV}$ &$E_{T}~\textgreater~160 ~\mathrm{GeV}$  &$E_{T}~\textgreater~220 ~\mathrm{GeV}$ \\	
\hline	
\end{tabular}
}
\end{table*}
%

\begin{table*}[!t]\scriptsize
	\centering{
		\caption{After different cuts applied, the cross sections for the signal and SM background at the $\sqrt{s}=$ 380 GeV CLIC $~~~~~~~~~~~~~~~~$for $M_{a}$ = 8, 10, 20, 100, 200 GeV and $g_{a\gamma\gamma}=10^{-3}~$ GeV$^{-1}$. \label{CLIC_table380}}
		\newcolumntype{C}[1]{>{\centering\let\newline\\\arraybackslash\hspace{0pt}}m{#1}}
		\begin{tabular}{|C{1.4cm}|C{1.57cm}|C{1.57cm}|C{1.57cm}|C{1.57cm}|C{1.57cm}|C{2.0cm}| }
			\hline
			\multicolumn{7}{|c|} { CLIC @ $\sqrt{s}=380$ GeV  }\\
			\hline
			\multirow{2}{*}{Cuts}    & \multicolumn{5}{c|}{Signal (fb) }&\multicolumn{1}{c|}{Background (fb) }   \\
			\cline{2-7}
			&$M_a$~=~8~GeV &$M_a$~=~10~GeV &$M_a$~=~20~GeV &$M_a$~=~100~GeV &$M_a$~= ~200~GeV&${\gamma \gamma \overline{\nu}{\nu}}$   \\
			\hline
			Basic cuts &0.3572 &0.3542 &0.3506 &0.3279 &0.0731   &12.66 \\
			Cut 1      &0.3436 &0.3416 &0.3364 &0.2762 &0.0612   &7.4612 \\
			Cut 2      &0.3404 &0.3384 &0.332  &0.2464 &0.0459   &5.5602 \\
			Cut 3      &0.3198 &0.318  &0.3118 &0.2442 &0.0457   &5.3916 \\
			\hline
	\end{tabular}}	
\end{table*}
\begin{table*}[!ht]\scriptsize
	\centering{
		\caption{Same as Table \ref{CLIC_table380} but for the $\sqrt{s}=$ 1500 GeV CLIC. \label{CLIC_table1500}}
		\newcolumntype{C}[1]{>{\centering\let\newline\\\arraybackslash\hspace{0pt}}m{#1}}
		\begin{tabular}{|C{1.4cm}|C{1.24cm}|C{1.24cm}|C{1.24cm}|C{1.24cm}|C{1.24cm}|C{1.24cm}|C{2.0cm}| }
			\hline
			\multicolumn{8}{|c|} { CLIC @ $\sqrt{s}=1500$ GeV  }\\
			\hline
			\multirow{2}{*}{Cuts}&\multicolumn{6}{c|}{Signal (fb) }&\multicolumn{1}{c|}{Background (fb) }   \\
			\cline{2-8}
			&$M_a$~= ~ 8~GeV   &$M_a$~= ~ 10~GeV&$M_a$~= ~20~GeV&$M_a$~= 100~GeV&$M_a$~= 500~GeV&$M_a$~= 1000~GeV&${\gamma \gamma \overline{\nu}{\nu}}$   \\
			\hline
			Basic cuts &6.6575 &8.8253 &12.176 &11.9803  &1.7862  &0.1573  &21.322 \\
			Cut 1      &3.7949 &5.4839 &8.6115 &8.3208  &1.0745  &0.0913  &8.1313 \\
			Cut 2      &3.3201 &4.9969 &8.12   &7.8021  &1.0991  &0.0756  &3.2893 \\
			Cut 3      &3.1913 &4.8613 &8.0028 &7.7769  &1.0089  &0.0756  &3.2211 \\
			\hline
		\end{tabular}	
	}
\end{table*}
\begin{table*}[!t]\scriptsize
	\centering{
		\caption{Same as Table \ref{CLIC_table380} but for the $\sqrt{s}=$ 3000 GeV CLIC. \label{CLIC_table3000}}
		\newcolumntype{C}[1]{>{\centering\let\newline\\\arraybackslash\hspace{0pt}}m{#1}}
		\begin{tabular}{|C{1.4cm}|C{1cm}|C{1cm}|C{1cm}|C{1cm}|C{1cm}|C{1cm}|C{1cm}|C{2.0cm}| }
			\hline
			\multicolumn{9}{|c|} { CLIC @ $\sqrt{s}=3000$ GeV  }\\
			\hline 
			\multirow{2}{*}{Cuts}    & \multicolumn{7}{c|}{Signal (fb) }&\multicolumn{1}{c|}{Background (fb) }   \\
			\cline{2-8}
			&$M_a$~= 8~GeV&$M_a$~=  10~GeV&$M_a$~=  20~GeV&$M_a$~= 100~GeV&$M_a$~= 1000~GeV&$M_a$~= 1500~GeV&$M_a$~= 2000~GeV&${\gamma \gamma \overline{\nu}{\nu}}$   \\
			\hline
			Basic cuts &9.6983    &14.4563  &27.993   &32.5997    &3.5915  &1.3519   &0.3323  &29.1063   \\
			Cut 1      &5.557     &8.7537   &20.1513  &24.8866    &2.2756  &0.8178   &0.2003  &14.56  \\
			Cut 2      &3.8733    &7.0693   &18.4306  &23.1416    &2.1683  &0.7647   &0.1769  &5.0493     \\
			Cut 3      &3.3481    &6.552    &17.903   &22.946    &2.1682  &0.7647   &0.1769  &4.8927    \\	
			\hline
		\end{tabular}
	}
\end{table*}

 We look for photons reconstructed in the barrel region $|\eta_{\gamma}| < 2.5$.
 A light ALP tends to decay a pair of collimated photons, which appear essentially as one photon of the combined energy in a detector and thus can not be correctly reconstructed as two individual photons~\cite{Sheff:2020jyw}. In Ref.~\cite{Jaeckel:2015jla}, the light ALP is studied via the decay $ Z\rightarrow \gamma a~(\to\gamma\gamma)$ with $\Delta R(\gamma,\gamma) \sim 4\frac{m_{a}}{m_{Z}}$. $\Delta R(\gamma,\gamma) = \sqrt{(\Delta \phi)^{2}+(\Delta \eta)^{2}}$ is the particle separation in the rapidity-azimuth plane, $\Delta \phi$ and $\Delta \eta$ are the azimuth difference and pseudo-rapidity difference of the photon pair, respectively. To reconstruct collimated pair of photons from the light ALP decay, a peak separation of $\Delta R(\gamma,\gamma) =4 \frac{m_{\pi}}{E_{\pi}} = 0.035$ is taken in Ref.~\cite{Steinberg:2021iay}, where $m_{\pi}$ and $E_{\pi}$  represent the mass and energy of $\pi^{0}$, and we can see that if $\Delta R(\gamma,\gamma)$ is smaller than 0.035, making photon pair reconstruction challenging. In our work, the peak of the normalized distribution of $\Delta R(\gamma,\gamma)$ is near 0.06 which allows for sufficient separation for both photons in the pair at $M_{a}=5$ GeV.
To better isolate produced photon pairs, we ask for the isolation between the two photons $\Delta R(\gamma,\gamma)>0.06$ and the photon having a minimum transverse momentum of 30 GeV. Based on the above analysis, the basic cuts are chosen as follows:
$$
\begin{array}{l}
	\left|\eta_{\gamma}\right|<2.5,  ~~ \quad 	\Delta R(\gamma, \gamma)> 0.06, ~~ \quad  p_{T}^{\gamma}>30~\mathrm{GeV}.\\
\end{array}
$$

In the following, the signal and background events are generated at tree-level by \textsf{Madgraph5-aMC@NLO}~\cite{Alwall:2014hca} with the above basic cuts.
The fast detector simulation are performed with \textsf{Delphes}~\cite{deFavereau:2013fsa} using the CLIC detector card. Finally, we use \textsf{MadAnalysis5}~\cite{Conte:2012fm} package to do the kinematic and cut-based analysis with the reconstructed-level events of signal and background. The final states of the signal and background contain invisible neutrino and two photons. For the signal, two photons in the final state from ALP decay could be a powerful trigger, and the angular separation of two photons depends largely on the ALP mass. Therefore, we choose the invariant mass of the photon pair $M (\gamma ~\gamma)$, the angle between reconstructed ALP and beam axis $\theta(\gamma~\gamma)$, the missing transverse energy $\slashed{E}_{T}$ associated to the invisible neutrino system, as well as the transverse energy $E_{T}$ associated to the visible particles as observables in the lab frame. The missing transverse energy and transverse energy are defined by $\slashed{E}_{T}= \arrowvert \sum\limits_{\rm visible~particles}\limits^{} \overrightarrow{p}_{T}\arrowvert $ and $E_{T}=\sum\limits_{\rm visible~particles}\limits^{}\arrowvert  \overrightarrow{p}_{T}\arrowvert $, where $\overrightarrow{p}_{T}$ stands for the visible particles transverse momentum.

In Fig.~\ref{CLIC_fig380}, we plot the normalized distributions of $M (\gamma ~\gamma)$, $\theta(\gamma~\gamma)$, $\slashed{E}_{T}$ and $E_{T}$ for the signal and background with the signal benchmark points $M_{a}$ = 8, 10, 20, 100, 200 GeV and the fixed parameter $g_{a\gamma\gamma}$~=~$10^{-3}$ GeV$^{-1}$ for the c.m. energy $\sqrt{s}$ $=$380 GeV.
 In Fig.~\ref{CLIC_fig1500} we show the same normalized kinematic distributions for $M_{a}$ = 8, 10, 20, 100, 500, 1000 GeV and $g_{a\gamma\gamma}=10^{-3}$ GeV$^{-1}$
 at $\sqrt{s}=$ 1500 GeV, and the same is shown in Fig.~\ref{CLIC_fig3000} for  $M_{a}$ = 8, 10, 20, 100, 1000, 1500, 2000 GeV and $g_{a\gamma\gamma}=10^{-3}$ GeV$^{-1}$ at  $\sqrt{s}=$ 3000 GeV.
In Figs. 4, 5 and 6, all the signal and background samples are obtained by after applying the above basic cuts.
 From Figs.~\ref{CLIC_fig380},~\ref{CLIC_fig1500} and~\ref{CLIC_fig3000}, we can see that, in order to optimize the signal significance, some additional cuts of kinematic distributions are needed.
 From these figures, we can see that it is not suitable to impose restrictions on the normalized $M(\gamma~\gamma)$ distribution, hence we do not put any constraint. 
 According to the behavioral characteristics of the other three distributions, we can further impose the following improved cuts on the signal and background events as shown in Table \ref{cuts}.
 Of course, we have applied the basic cuts before applying the cuts in Table \ref{cuts}, and the cuts 1, 2 and 3 are applied in sequence.

After imposing the cuts for a few representative ALP mass benchmark points, we summarize the cross sections of the signal and background in Tables~\ref{CLIC_table380},~\ref{CLIC_table1500} and~\ref{CLIC_table3000} for $\sqrt{s}=$ 380 GeV, 1500 GeV and 3000 GeV, respectively.
From Tables~\ref{CLIC_table380},~\ref{CLIC_table1500} and~\ref{CLIC_table3000}, after applying the basic cuts we can find that Tables~\ref{CLIC_table1500} and~\ref{CLIC_table3000} show very large differences in the cross sections between $M_a=8, 10$ and $20$ GeV, while the variations appear to be minimal in Table~\ref{CLIC_table380}. For the ALP-photon coupling $g_{a\gamma\gamma}$~=~$10^{-3}$ GeV$^{-1}$, the variations in the cross section of signal are minimal in the ALP mass interval 5 GeV $\sim$ 20 GeV without cut applied at $\sqrt{s} =$ 380, 1500 and 3000  GeV. The peak of the distribution shifts towards smaller $\Delta R (\gamma, \gamma)$ for the larger c.m. energy at same ALP mass, while the range of $|\eta_{\gamma}|$ for the high c.m. energy is larger than that for the low c.m. energy at same ALP mass.
That is to say, for $\sqrt{s} =$ 3 TeV, the signal for $M_a =$ 8, 10 and 20 GeV lie very close to the signal region boundaries, so the lighter the ALP the more the signal is cut out.
Therefore, the effects of the basic cuts on the cross sections of $M_a=8, 10$ and $20$ GeV at $\sqrt{s}=1500$ and $ 3000 $ GeV are greater than those at $\sqrt{s}=380 $ GeV.
From these tables, one can see that  the background is suppressed very efficiently, while the signal still has good efficiency after imposing all cuts.

We use the following Poisson formula~\cite{Cowan:2010js} to estimate the statistical significance (SS) at the CLIC with the luminosities of 0.5 ab$^{-1}$, 1.5 ab$^{-1}$ and 3 ab$^{-1}$, respectively,
\begin{equation}\label{SS}
\quad \quad	SS = \sqrt{2 \mathcal{L} [(S + B)\ln(1+\frac{S}{B})-S]},
\end{equation}
where $\mathcal{L}$ represents the integrated luminosity, $S$ and $B$ respectively represent the effective cross sections of the signal and background after imposing all cuts.

In Fig.~\ref{3-5sigmavs}, we present the 3$\sigma$ and 5$\sigma$ bounds on the ALP parameter space from the W$^{+}$W$^{-}$ fusion process e$^{+}$e$^{-}$ $\rightarrow$ $\overline{\nu}$$_{e}$$\nu$$_{e}a$ $(\rightarrow \gamma \gamma)$ at the CLIC. The blue, red and green lines depict the bounds coming from the analysis for the three energy stages of the CLIC.
This figure shows that the sensitivity bounds on the ALP-photon coupling $g_{a\gamma\gamma}$ can be improved to 0.8 TeV${^{-1}}$ (1.1 TeV${^{-1}}$) with the first energy stage, 0.13 TeV${^{-1}}$ (0.15 TeV${^{-1}}$) with the second energy stage and 0.073 TeV${^{-1}}$ (0.091 TeV${^{-1}}$) with the third energy stage for the ALP mass interval 5 GeV $\sim$ 340 GeV, 5 GeV $\sim$ 1300 GeV and 5 GeV $\sim$ 2600 GeV at 3$\sigma$ (5$\sigma$) levels, respectively.  Comparing our results of the three energy stages at the CLIC with each other, it is obvious that the sensitivity bound increases with the c.m. energy increases.
 
Our obtained bounds  and other current existing constraints on the ALP-photon coupling are presented in Fig.~\ref{2sigmavs}. The green region comes from different beam dump experiments\\ \cite{Riordan:1987aw,Bjorken:1988as,Dobrich:2015jyk}.
The dark blue and light blue regions depict the results given by ~\cite{Mimasu:2014nea,Jaeckel:2015jla} via the diphoton and triphoton final states (e$^{+}$e$^{-}~\to~2\gamma,~3\gamma$) at the LEP. The dark gray and black regions denote  the constraints from radiative $\Upsilon~\to~\gamma a$ at the Babar~\cite{BaBar:2011kau} and the diphoton resonance searches at the LHCb~\cite{CidVidal:2018blh}, respectively.
The dark green region labelled as "L3"  represents looking for isolated and energetic photons from hadronic decay of Z boson at the L3 collaboration~\cite{L3:1992kcg}. The exclusion regions also include the results from the LHC resonant $\gamma\gamma$ searches in red~\cite{Jaeckel:2012yz,ATLAS:2012fgo,Mariotti:2017vtv}.
Our projected CLIC sensitivity for $\sqrt{s}=$ 380, 1500 and 3000 GeV are enclosed by the cyan, blue and wine lines, respectively.
 Comparing our exclusion limits with the results of the LEP and LHC, we can conjecture that the CLIC has better potential to search for ALPs in the lower ALP mass interval 5 GeV $\sim$ 50 GeV and high ALP mass interval 2 TeV $\sim$ 2.7 TeV.

\section{Conclusions }\label{conclusions}

\indent

As a class of attractive particles, ALPs may influence the structure Electroweak phase transition, play an important role in solving the hierarchy problem and serve as cold dark matter (DM). There are excellent motivations to study ALPs. In this paper, we study the CLIC potential to search for ALP via the W$^{+}$W$^{-}$ fusion  process e$^{+}$e$^{-}$ $\rightarrow$ $\overline{\nu}$$_{e}$$\nu$$_{e}a$ $(\rightarrow \gamma \gamma)$.
We have performed the numerical calculations and the phenomenological analysis for the signal and relevant SM background and obtained the 3$\sigma$ and 5$\sigma$ bounds on the ALP parameter space at the three energy stages of CLIC. Comparing our numerical results with other existing bounds, we find that, for the low mass region 5 GeV $\sim$ 50 GeV, the constraint on the ALP-photon coupling $g_{a\gamma\gamma}$  can be improved to 5$\times$10$^{-2}$ TeV$^{-1}$. Meanwhile, our result complements the research of ALP-photon coupling by the ATLAS and CMS in the ALP mass range of 2 TeV $\sim$ 2.7 TeV.

In conclusion, we have found that searching for ALPs at the CLIC via the W$^{+}$W$^{-}$ fusion process e$^{+}$e$^{-}$ $\rightarrow$  $\overline{\nu}$$_{e}$$\nu$$_{e}a$ $(\rightarrow \gamma \gamma)$ can not only cover low mass ALPs, but also have the greater potential to study the high mass ALPs. Thus, studying ALPs through the W$^{+}$W$^{-}$ fusion process e$^{+}$e$^{-}$ $\rightarrow$~$\overline{\nu}$$_{e}$$\nu$$_{e}a$ $(\rightarrow \gamma \gamma)$  at the CLIC provides a complementary window to the opportunities offered by other high energy collider experiments.
\section*{Acknowledgments}
This work was supported in part by the National Natural Science Foundation of China under Grants No. 11875157 and 12147214.

\bibliography{References}

\begin{thebibliography}{10}

\bibitem{Aad:2012tfa}
Georges Aad et~al.
\newblock {Observation of a new particle in the search for the Standard Model
  Higgs boson with the ATLAS detector at the LHC}.
\newblock {\em Phys. Lett. B}, 716:1--29, 2012.

\bibitem{Chatrchyan:2012ufa}
Serguei Chatrchyan et~al.
\newblock {Observation of a New Boson at a Mass of 125 GeV with the CMS
  Experiment at the LHC}.
\newblock {\em Phys. Lett. B}, 716:30--61, 2012.

\bibitem{Djouadi:2007ik}
Gerald Aarons et~al.
\newblock {International Linear Collider Reference Design Report Volume 2:
  Physics at the ILC}.
\newblock 9 2007.

\bibitem{Baer:2013cma}
{The International Linear Collider Technical Design Report - Volume 2:
  Physics}.
\newblock 6 2013.

\bibitem{Moortgat-Picka:2015yla}
A.~Arbey et~al.
\newblock {Physics at the e+ e- Linear Collider}.
\newblock {\em Eur. Phys. J. C}, 75(8):371, 2015.

\bibitem{Aihara:2019gcq}
Hiroaki Aihara et~al.
\newblock {The International Linear Collider. A Global Project}.
\newblock 1 2019.

\bibitem{Bambade:2019fyw}
Philip Bambade et~al.
\newblock {The International Linear Collider: A Global Project}.
\newblock 3 2019.

\bibitem{Battaglia:2004mw}
E.~Accomando et~al.
\newblock {Physics at the CLIC multi-TeV linear collider}.
\newblock In {\em {11th International Conference on Hadron Spectroscopy}}, CERN
  Yellow Reports: Monographs, 6 2004.

\bibitem{Roloff:2018dqu}
{The Compact Linear e$^+$e$^-$ Collider (CLIC): Physics Potential}.
\newblock 12 2018.

\bibitem{Gomez-Ceballos:2013zzn}
M.~Bicer et~al.
\newblock {First Look at the Physics Case of TLEP}.
\newblock {\em JHEP}, 01:164, 2014.

\bibitem{Abada:2019zxq}
A.~Abada et~al.
\newblock {FCC-ee: The Lepton Collider}: {Future Circular Collider Conceptual
  Design Report Volume 2}.
\newblock {\em Eur. Phys. J. ST}, 228(2):261--623, 2019.

\bibitem{CEPCStudyGroup:2018rmc}
{CEPC Conceptual Design Report: Volume 1 - Accelerator}.
\newblock 9 2018.

\bibitem{CEPCStudyGroup:2018ghi}
Mingyi Dong et~al.
\newblock {CEPC Conceptual Design Report: Volume 2 - Physics \& Detector}.
\newblock 11 2018.

\bibitem{An:2018dwb}
Fenfen An et~al.
\newblock {Precision Higgs physics at the CEPC}.
\newblock {\em Chin. Phys. C}, 43(4):043002, 2019.

\bibitem{Peccei:1977hh}
R.~D. Peccei and Helen~R. Quinn.
\newblock {CP Conservation in the Presence of Instantons}.
\newblock {\em Phys. Rev. Lett.}, 38:1440--1443, 1977.

\bibitem{Weinberg:1977ma}
Steven Weinberg.
\newblock {A New Light Boson?}
\newblock {\em Phys. Rev. Lett.}, 40:223--226, 1978.

\bibitem{Wilczek:1977pj}
Frank Wilczek.
\newblock {Problem of Strong $P$ and $T$ Invariance in the Presence of
  Instantons}.
\newblock {\em Phys. Rev. Lett.}, 40:279--282, 1978.

\bibitem{Kim:1979if}
Jihn~E. Kim.
\newblock {Weak Interaction Singlet and Strong CP Invariance}.
\newblock {\em Phys. Rev. Lett.}, 43:103, 1979.

\bibitem{Shifman:1979if}
Mikhail~A. Shifman, A.~I. Vainshtein, and Valentin~I. Zakharov.
\newblock {Can Confinement Ensure Natural CP Invariance of Strong
  Interactions?}
\newblock {\em Nucl. Phys. B}, 166:493--506, 1980.

\bibitem{Dine:1981rt}
Michael Dine, Willy Fischler, and Mark Srednicki.
\newblock {A Simple Solution to the Strong CP Problem with a Harmless Axion}.
\newblock {\em Phys. Lett. B}, 104:199--202, 1981.

\bibitem{Zhitnitsky:1980tq}
A.~R. Zhitnitsky.
\newblock {On Possible Suppression of the Axion Hadron Interactions. (In
  Russian)}.
\newblock {\em Sov. J. Nucl. Phys.}, 31:260, 1980.

\bibitem{Bauer:2017ris}
Martin Bauer, Matthias Neubert, and Andrea Thamm.
\newblock {Collider Probes of Axion-Like Particles}.
\newblock {\em JHEP}, 12:044, 2017.

\bibitem{Bauer:2018uxu}
Martin Bauer, Mathias Heiles, Matthias Neubert, and Andrea Thamm.
\newblock {Axion-Like Particles at Future Colliders}.
\newblock {\em Eur. Phys. J. C}, 79(1):74, 2019.

\bibitem{Fortin:2021cog}
Jean-Fran\c{c}ois Fortin, Huai-Ke Guo, Steven~P. Harris, Doojin Kim, Kuver
  Sinha, and Chen Sun.
\newblock {Axions: From magnetars and neutron star mergers to beam dumps and
  BECs}.
\newblock {\em Int. J. Mod. Phys. D}, 30(07):2130002, 2021.

\bibitem{dEnterria:2021ljz}
David d'Enterria.
\newblock {Collider constraints on axion-like particles}.
\newblock In {\em {Workshop on Feebly Interacting Particles}}, 2 2021.

\bibitem{Agrawal:2021dbo}
Prateek Agrawal et~al.
\newblock {Feebly-interacting particles: FIPs 2020 workshop report}.
\newblock {\em Eur. Phys. J. C}, 81(11):1015, 2021.

\bibitem{Brivio:2017ije}
I.~Brivio, M.~B. Gavela, L.~Merlo, K.~Mimasu, J.~M. No, R.~del Rey, and
  V.~Sanz.
\newblock {ALPs Effective Field Theory and Collider Signatures}.
\newblock {\em Eur. Phys. J. C}, 77(8):572, 2017.

\bibitem{Knapen:2016moh}
Simon Knapen, Tongyan Lin, Hou~Keong Lou, and Tom Melia.
\newblock {Searching for Axionlike Particles with Ultraperipheral Heavy-Ion
  Collisions}.
\newblock {\em Phys. Rev. Lett.}, 118(17):171801, 2017.

\bibitem{Baldenegro:2019whq}
C.~Baldenegro, S.~Hassani, C.~Royon, and L.~Schoeffel.
\newblock {Extending the constraint for axion-like particles as resonances at
  the LHC and laser beam experiments}.
\newblock {\em Phys. Lett. B}, 795:339--345, 2019.

\bibitem{Coelho:2020saz}
R.~O. Coelho, V.~P. Goncalves, D.~E. Martins, and M.~S. Rangel.
\newblock {Production of axionlike particles in $PbPb$ collisions at the LHC,
  HE\textendash{}LHC and FCC: A phenomenological analysis}.
\newblock {\em Phys. Lett. B}, 806:135512, 2020.

\bibitem{Coelho:2020syp}
R.~O. Coelho, V.~P. Gon\c{c}alves, D.~E. Martins, and M.~Rangel.
\newblock {Exclusive and diffractive $\gamma \gamma$ production in $PbPb$
  collisions at the LHC, HE - LHC and FCC}.
\newblock {\em Eur. Phys. J. C}, 80(5):488, 2020.

\bibitem{Goncalves:2021pdc}
V.~P. Goncalves, D.~E. Martins, and M.~S. Rangel.
\newblock {Searching for axionlike particles with low masses in pPb and PbPb
  collisions}.
\newblock {\em Eur. Phys. J. C}, 81(6):522, 2021.

\bibitem{Jaeckel:2012yz}
Joerg Jaeckel, Martin Jankowiak, and Michael Spannowsky.
\newblock {LHC probes the hidden sector}.
\newblock {\em Phys. Dark Univ.}, 2:111--117, 2013.

\bibitem{Baldenegro:2018hng}
Cristian Baldenegro, Sylvain Fichet, Gero von Gersdorff, and Christophe Royon.
\newblock {Searching for axion-like particles with proton tagging at the LHC}.
\newblock {\em JHEP}, 06:131, 2018.

\bibitem{Florez:2021zoo}
Andr\'es Fl\'orez, Alfredo Gurrola, Will Johns, Paul Sheldon, Elijah Sheridan,
  Kuver Sinha, and Brandon Soubasis.
\newblock {Probing axionlike particles with $\gamma\gamma$ final states from
  vector boson fusion processes at the LHC}.
\newblock {\em Phys. Rev. D}, 103(9):095001, 2021.

\bibitem{Marciano:2016yhf}
W.~J. Marciano, A.~Masiero, P.~Paradisi, and M.~Passera.
\newblock {Contributions of axionlike particles to lepton dipole moments}.
\newblock {\em Phys. Rev. D}, 94(11):115033, 2016.

\bibitem{Inan:2020aal}
S.~C. \.Inan and A.~V. Kisselev.
\newblock {A search for axion-like particles in light-by-light scattering at
  the CLIC}.
\newblock {\em JHEP}, 06:183, 2020.

\bibitem{Inan:2020kif}
S.~C. \.Inan and A.~V. Kisselev.
\newblock {Polarized light-by-light scattering at the CLIC induced by
  axion-like particles}.
\newblock {\em Chin. Phys. C}, 45(4):043109, 2021.

\bibitem{Zhang:2021sio}
Hua-Ying Zhang, Chong-Xing Yue, Yu-Chen Guo, and Shuo Yang.
\newblock {Searching for axionlike particles at future electron-positron
  colliders}.
\newblock {\em Phys. Rev. D}, 104(9):096008, 2021.

\bibitem{Buttazzo:2018qqp}
Dario Buttazzo, Diego Redigolo, Filippo Sala, and Andrea Tesi.
\newblock {Fusing Vectors into Scalars at High Energy Lepton Colliders}.
\newblock {\em JHEP}, 11:144, 2018.

\bibitem{Gavela:2019cmq}
M.~B. Gavela, J.~M. No, V.~Sanz, and J.~F. de~Troc\'oniz.
\newblock {Nonresonant Searches for Axionlike Particles at the LHC}.
\newblock {\em Phys. Rev. Lett.}, 124(5):051802, 2020.

\bibitem{Georgi:1986df}
Howard Georgi, David~B. Kaplan, and Lisa Randall.
\newblock {Manifesting the Invisible Axion at Low-energies}.
\newblock {\em Phys. Lett. B}, 169:73--78, 1986.

\bibitem{Wang:2021uyb}
Daohan Wang, Lei Wu, Jin~Min Yang, and Mengchao Zhang.
\newblock {Photon-jet events as a probe of axionlike particles at the LHC}.
\newblock {\em Phys. Rev. D}, 104(9):095016, 2021.

\bibitem{Alloul:2013bka}
Adam Alloul, Neil~D. Christensen, C\'eline Degrande, Claude Duhr, and Benjamin
  Fuks.
\newblock {FeynRules 2.0 - A complete toolbox for tree-level phenomenology}.
\newblock {\em Comput. Phys. Commun.}, 185:2250--2300, 2014.

\bibitem{Azzurri:2021nmy}
Paolo Azzurri, Gregorio Bernardi, Sylvie Braibant, David d'Enterria, Jan
  Eysermans, Patrick Janot, Ang Li, and Emmanuel Perez.
\newblock {A special Higgs challenge: measuring the mass and production cross
  section with ultimate precision at FCC-ee}.
\newblock {\em Eur. Phys. J. Plus}, 137(1):23, 2022.

\bibitem{Jaeckel:2015jla}
Joerg Jaeckel and Michael Spannowsky.
\newblock {Probing MeV to 90 GeV axion-like particles with LEP and LHC}.
\newblock {\em Phys. Lett. B}, 753:482--487, 2016.

\bibitem{Mimasu:2014nea}
Ken Mimasu and Ver\'onica Sanz.
\newblock {ALPs at Colliders}.
\newblock {\em JHEP}, 06:173, 2015.

\bibitem{CLIC:2016zwp}
M~J Boland et~al.
\newblock {Updated baseline for a staged Compact Linear Collider}.
\newblock 8 2016.

\bibitem{Sheff:2020jyw}
Benjamin Sheff, Noah Steinberg, and James~D. Wells.
\newblock {Higgs boson decays into narrow diphoton jets and their search
  strategies at the Large Hadron Collider}.
\newblock {\em Phys. Rev. D}, 104(3):036009, 2021.

\bibitem{Steinberg:2021iay}
Noah Steinberg and James~D. Wells.
\newblock {Axion-Like Particles at the ILC Giga-Z}.
\newblock {\em JHEP}, 08:120, 2021.

\bibitem{Alwall:2014hca}
J.~Alwall, R.~Frederix, S.~Frixione, V.~Hirschi, F.~Maltoni, O.~Mattelaer,
  H.~S. Shao, T.~Stelzer, P.~Torrielli, and M.~Zaro.
\newblock {The automated computation of tree-level and next-to-leading order
  differential cross sections, and their matching to parton shower
  simulations}.
\newblock {\em JHEP}, 07:079, 2014.

\bibitem{deFavereau:2013fsa}
J.~de~Favereau, C.~Delaere, P.~Demin, A.~Giammanco, V.~Lema\^\i{}tre,
  A.~Mertens, and M.~Selvaggi.
\newblock {DELPHES 3, A modular framework for fast simulation of a generic
  collider experiment}.
\newblock {\em JHEP}, 02:057, 2014.

\bibitem{Conte:2012fm}
Eric Conte, Benjamin Fuks, and Guillaume Serret.
\newblock {MadAnalysis 5, A User-Friendly Framework for Collider
  Phenomenology}.
\newblock {\em Comput. Phys. Commun.}, 184:222--256, 2013.

\bibitem{Cowan:2010js}
Glen Cowan, Kyle Cranmer, Eilam Gross, and Ofer Vitells.
\newblock {Asymptotic formulae for likelihood-based tests of new physics}.
\newblock {\em Eur. Phys. J. C}, 71:1554, 2011.
\newblock [Erratum: Eur.Phys.J.C 73, 2501 (2013)].

\bibitem{Riordan:1987aw}
E.~M. Riordan et~al.
\newblock {A Search for Short Lived Axions in an Electron Beam Dump
  Experiment}.
\newblock {\em Phys. Rev. Lett.}, 59:755, 1987.

\bibitem{Bjorken:1988as}
J.~D. Bjorken, S.~Ecklund, W.~R. Nelson, A.~Abashian, C.~Church, B.~Lu, L.~W.
  Mo, T.~A. Nunamaker, and P.~Rassmann.
\newblock {Search for Neutral Metastable Penetrating Particles Produced in the
  SLAC Beam Dump}.
\newblock {\em Phys. Rev. D}, 38:3375, 1988.

\bibitem{Dobrich:2015jyk}
Babette D\"obrich, Joerg Jaeckel, Felix Kahlhoefer, Andreas Ringwald, and Kai
  Schmidt-Hoberg.
\newblock {ALPtraum: ALP production in proton beam dump experiments}.
\newblock {\em JHEP}, 02:018, 2016.

\bibitem{BaBar:2011kau}
J.~P. Lees et~al.
\newblock {Search for hadronic decays of a light Higgs boson in the radiative
  decay $\Upsilon \to \gamma A^0$}.
\newblock {\em Phys. Rev. Lett.}, 107:221803, 2011.

\bibitem{CidVidal:2018blh}
Xabier Cid~Vidal, Alberto Mariotti, Diego Redigolo, Filippo Sala, and Kohsaku
  Tobioka.
\newblock {New Axion Searches at Flavor Factories}.
\newblock {\em JHEP}, 01:113, 2019.
\newblock [Erratum: JHEP 06, 141 (2020)].

\bibitem{L3:1992kcg}
O.~Adriani et~al.
\newblock {Isolated hard photon emission in hadronic Z0 decays}.
\newblock {\em Phys. Lett. B}, 292:472--484, 1992.

\bibitem{ATLAS:2012fgo}
Georges Aad et~al.
\newblock {Measurement of isolated-photon pair production in $pp$ collisions at
  $\sqrt{s}=7$ TeV with the ATLAS detector}.
\newblock {\em JHEP}, 01:086, 2013.

\bibitem{Mariotti:2017vtv}
Alberto Mariotti, Diego Redigolo, Filippo Sala, and Kohsaku Tobioka.
\newblock {New LHC bound on low-mass diphoton resonances}.
\newblock {\em Phys. Lett. B}, 783:13--18, 2018.

\end{thebibliography}
\balance
\bibliographystyle{unsrt}
 
\begin{onecolumn}
\begin{appendices}
	
\section*{Appendix} \label{Appendix}

\vspace{20mm}
\begin{figure*}[!h]
	\centering{
		\includegraphics [width=15cm,height=4cm] {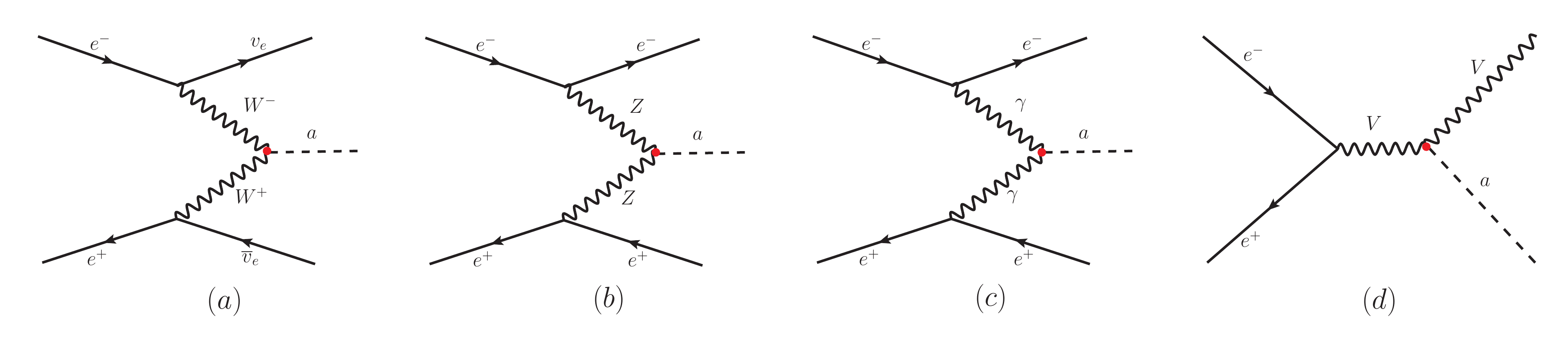}
		\caption{\label{feynman-4} The Feynman diagrams of the single production of ALP with V = $\gamma$, Z at e$^{+}$e$^{-}$ colliders.}}
\end{figure*}

\vspace{20mm}

\begin{figure*}[!h]	
	\centering
	\begin{subfigure}{0.37\linewidth}
		\includegraphics[width=\linewidth]{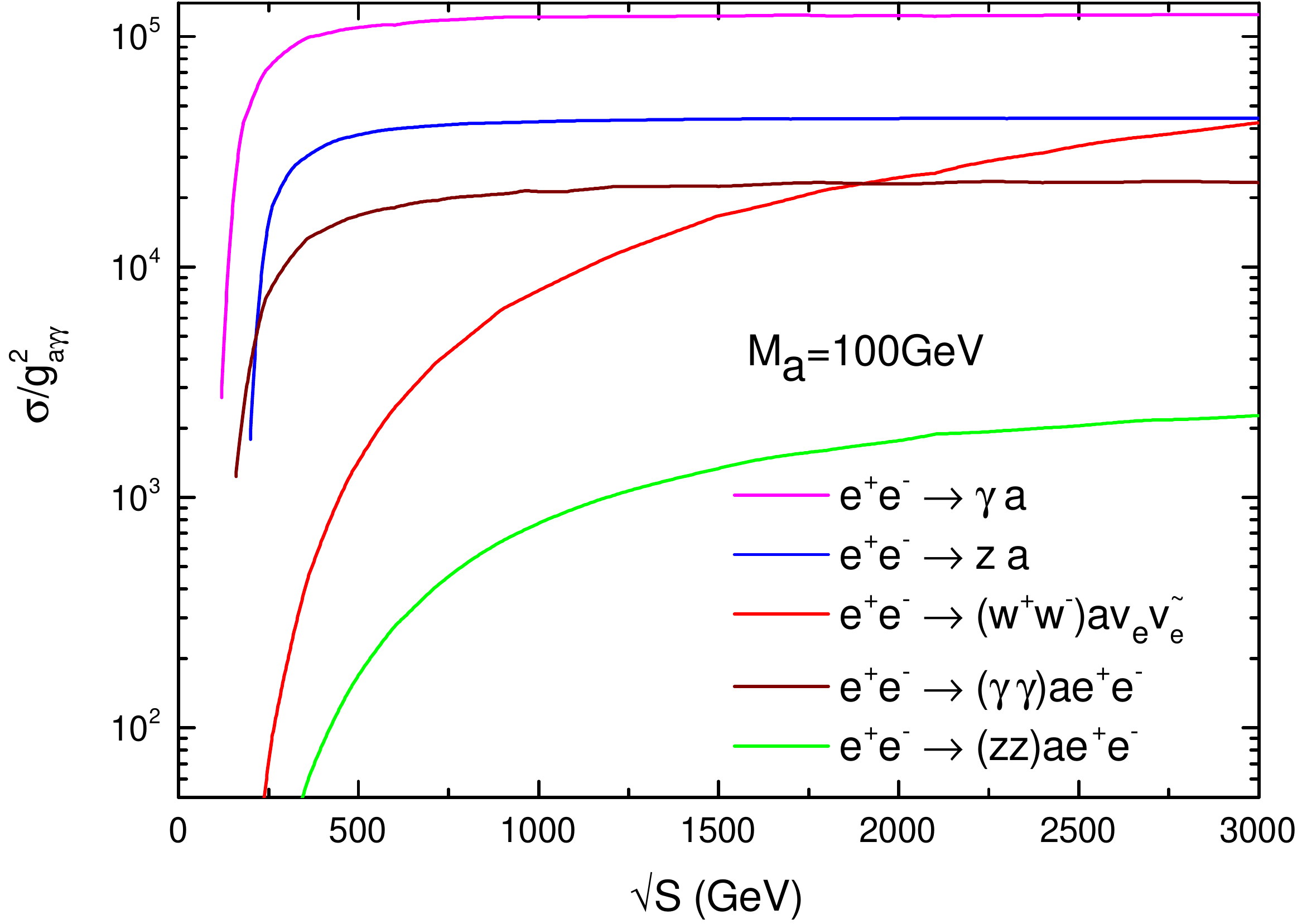}
	\end{subfigure}
	\begin{subfigure}{0.37\linewidth}
		\includegraphics[width=\linewidth]{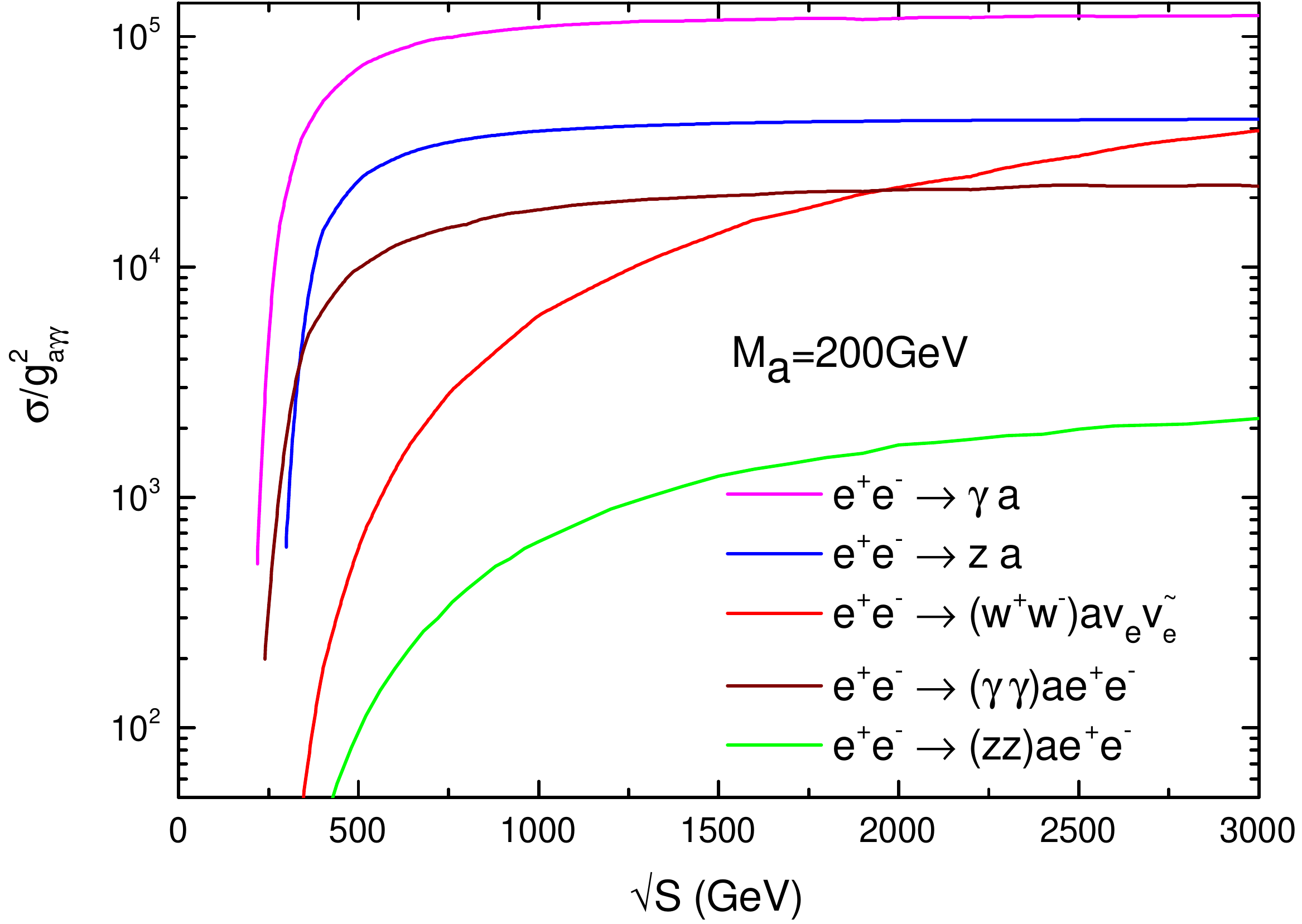}
	\end{subfigure}
	\begin{subfigure}{0.37\linewidth}
		\includegraphics[width=\linewidth]{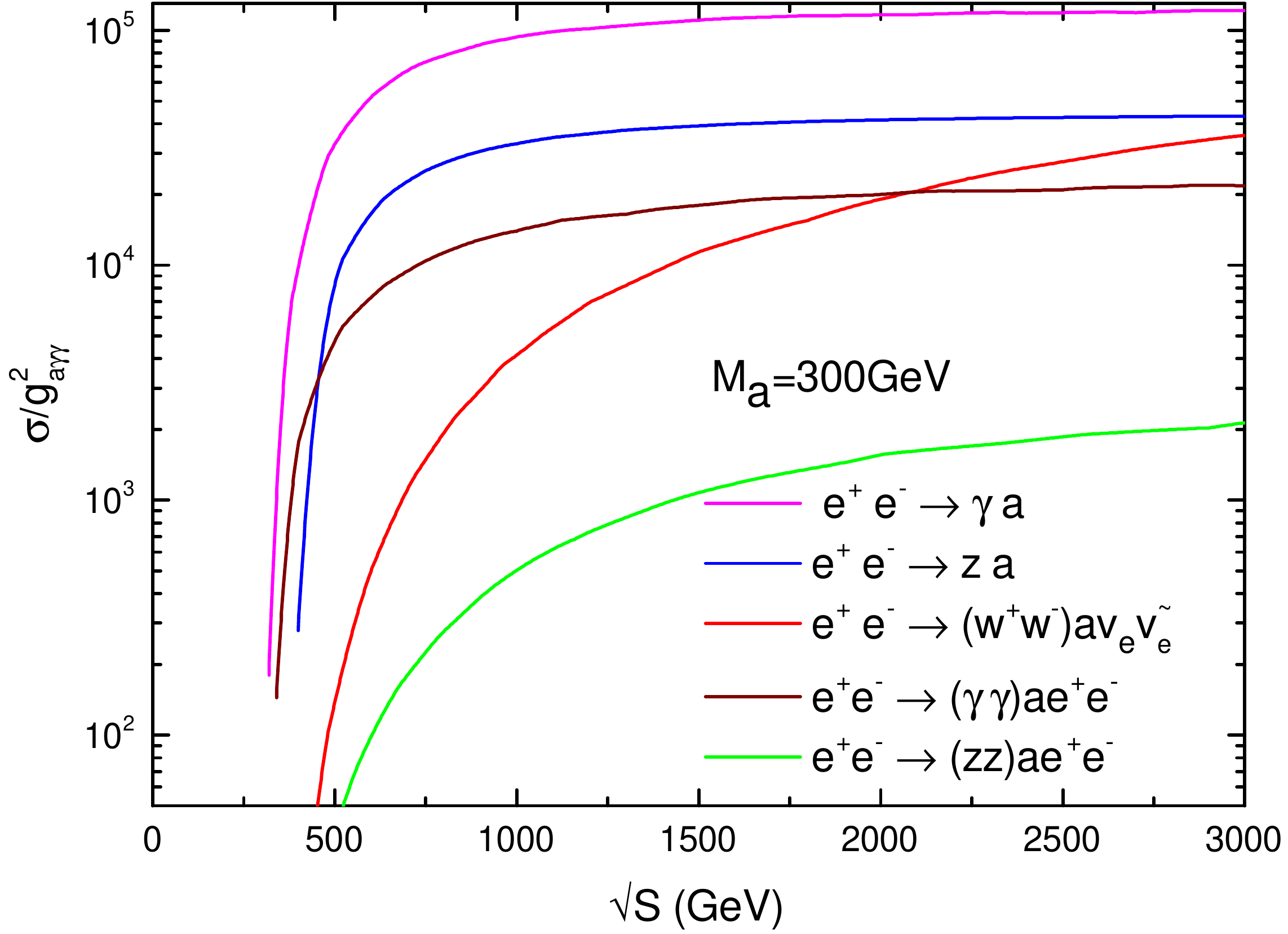}
	\end{subfigure}
	\begin{subfigure}{0.37\linewidth}
		\includegraphics[width=\linewidth]{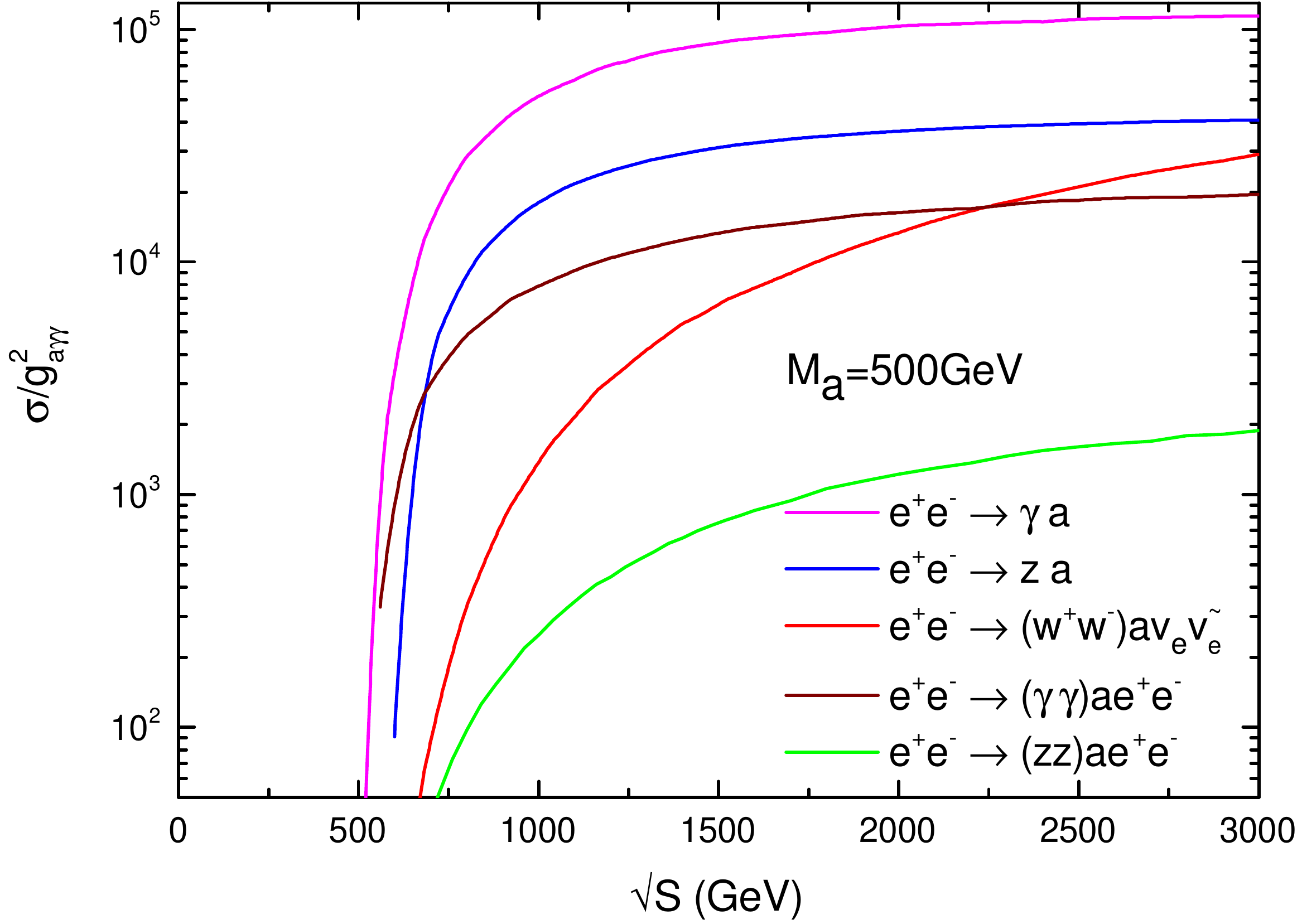}
	\end{subfigure}
	\caption{The ratio $\sigma$/$(g_{a\gamma\gamma})^{2}$ as a function of the c.m. energy $\sqrt{s}$ for $M_a$ = 100, 200, 300 and 500 GeV. }
	\label{sigmavsga}
\end{figure*}

\newpage
 \vspace{5mm}

\begin{figure*}[h]	
	\centering
	\includegraphics [height=7cm]  {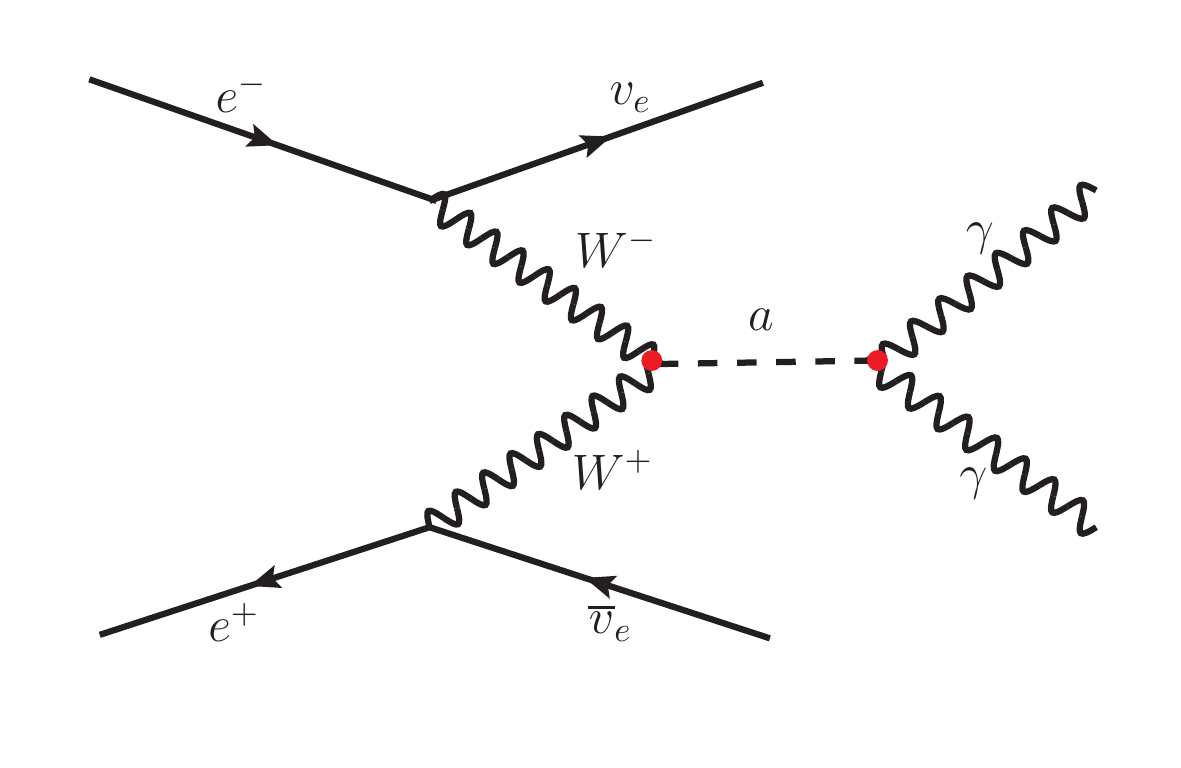}	
	\caption{Feynman diagram of the W$^{+}$W$^{-}$ fusion process e$^{+}$e$^{-}$ $\rightarrow$ $\overline{\nu}_{e}$$\nu_{e}$$a$ ($\to\gamma\gamma$).
		\label{signal-alp-decay}}
\end{figure*}

\vspace{20mm}

\begin{figure*}[h]	
	\centering
	\begin{subfigure}{0.37\linewidth}
		\includegraphics[width=\linewidth]{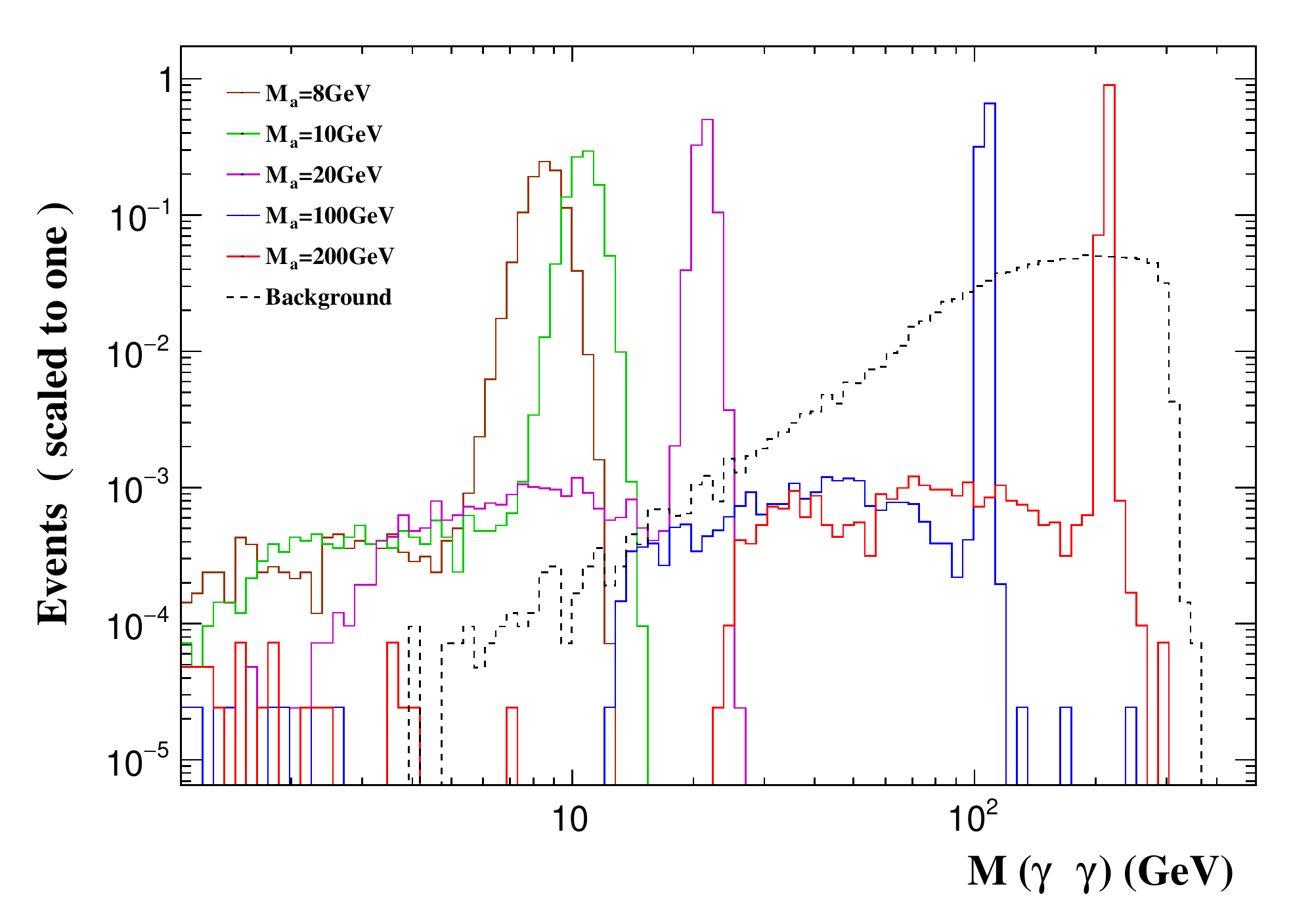}
	\end{subfigure}
	\begin{subfigure}{0.37\linewidth}
		\includegraphics[width=\linewidth]{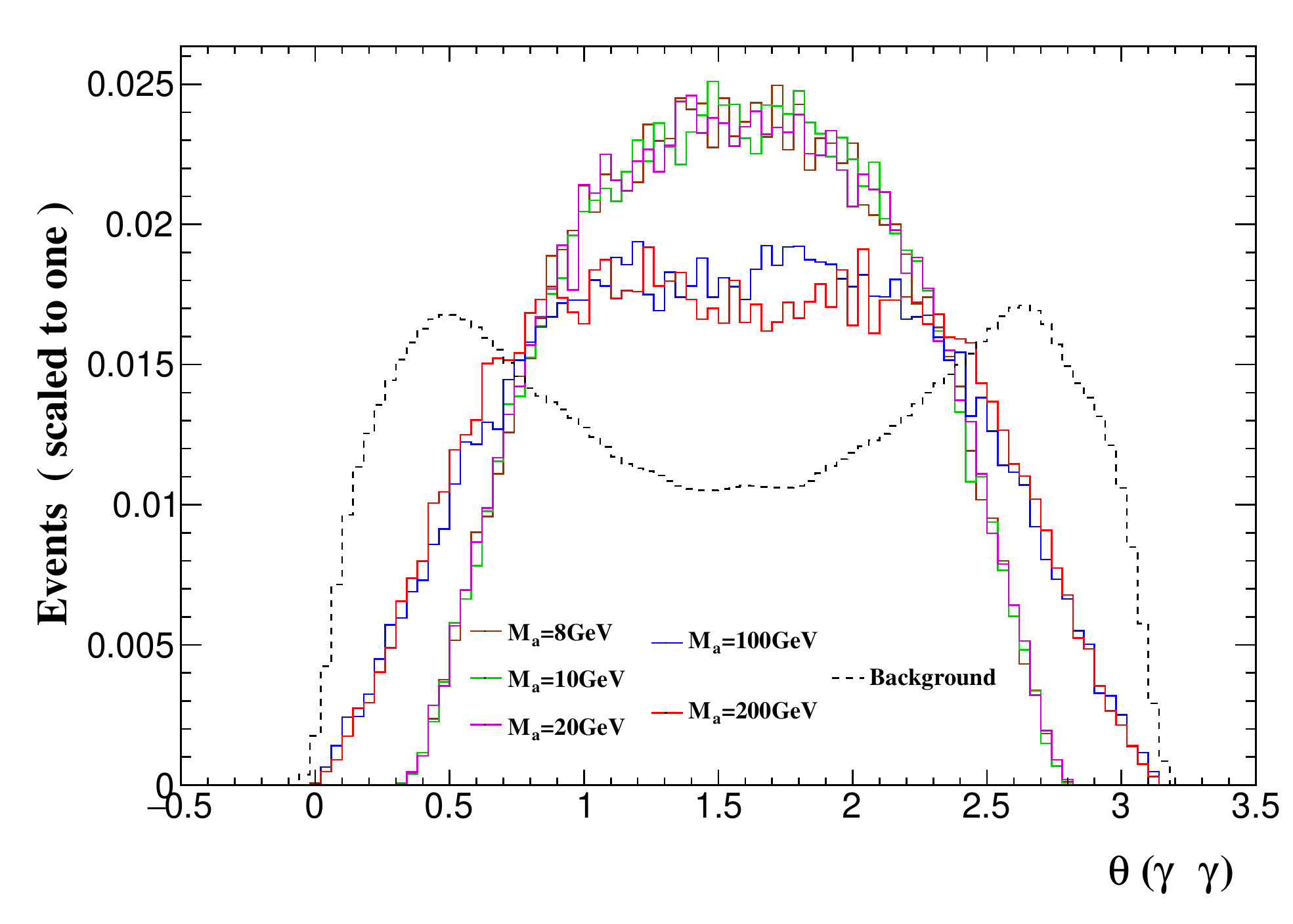}
	\end{subfigure}
	\begin{subfigure}{0.37 \linewidth}
		\includegraphics[width=\linewidth]{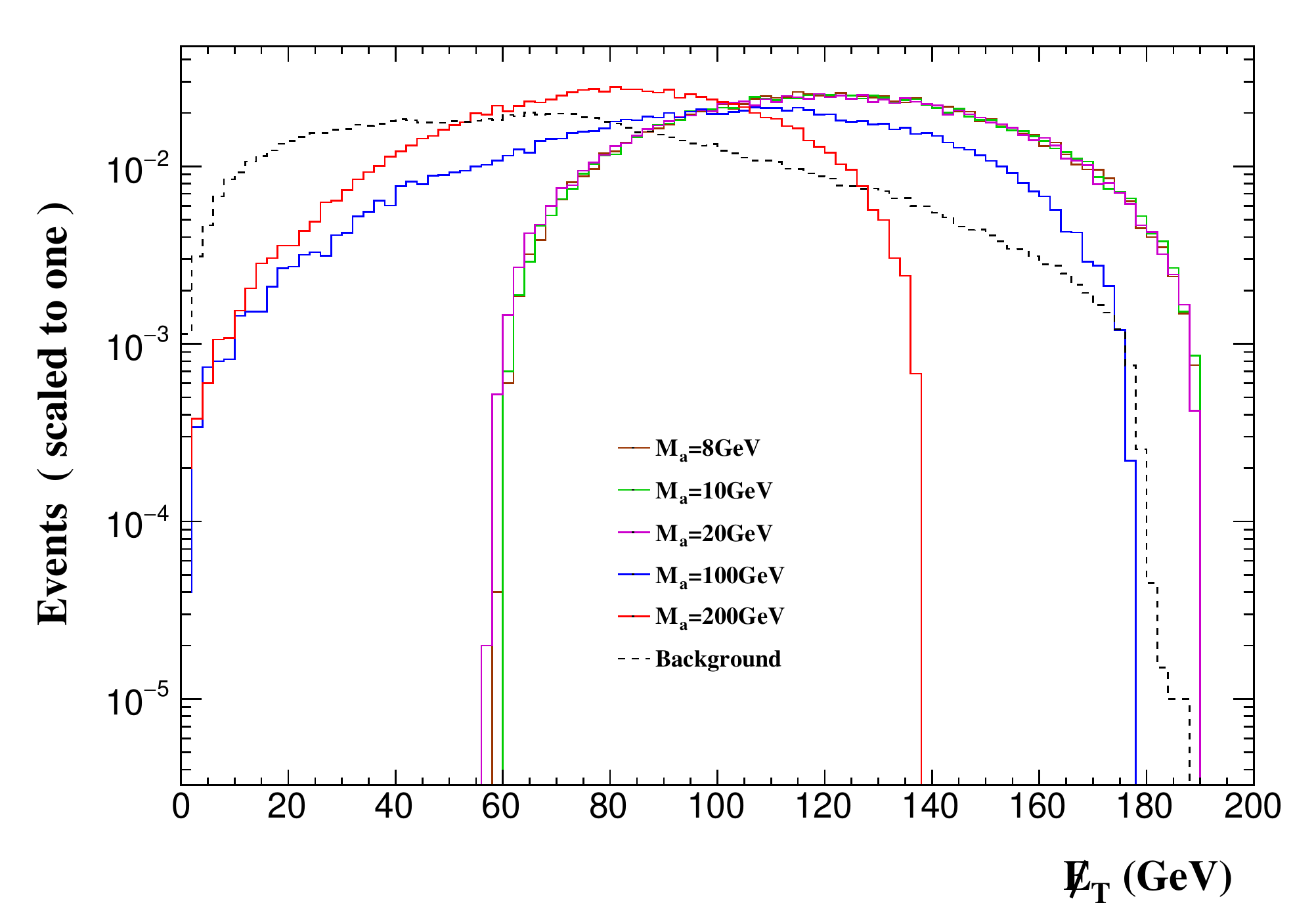}
	\end{subfigure}
	\begin{subfigure}{0.37 \linewidth}
		\includegraphics[width=\linewidth]{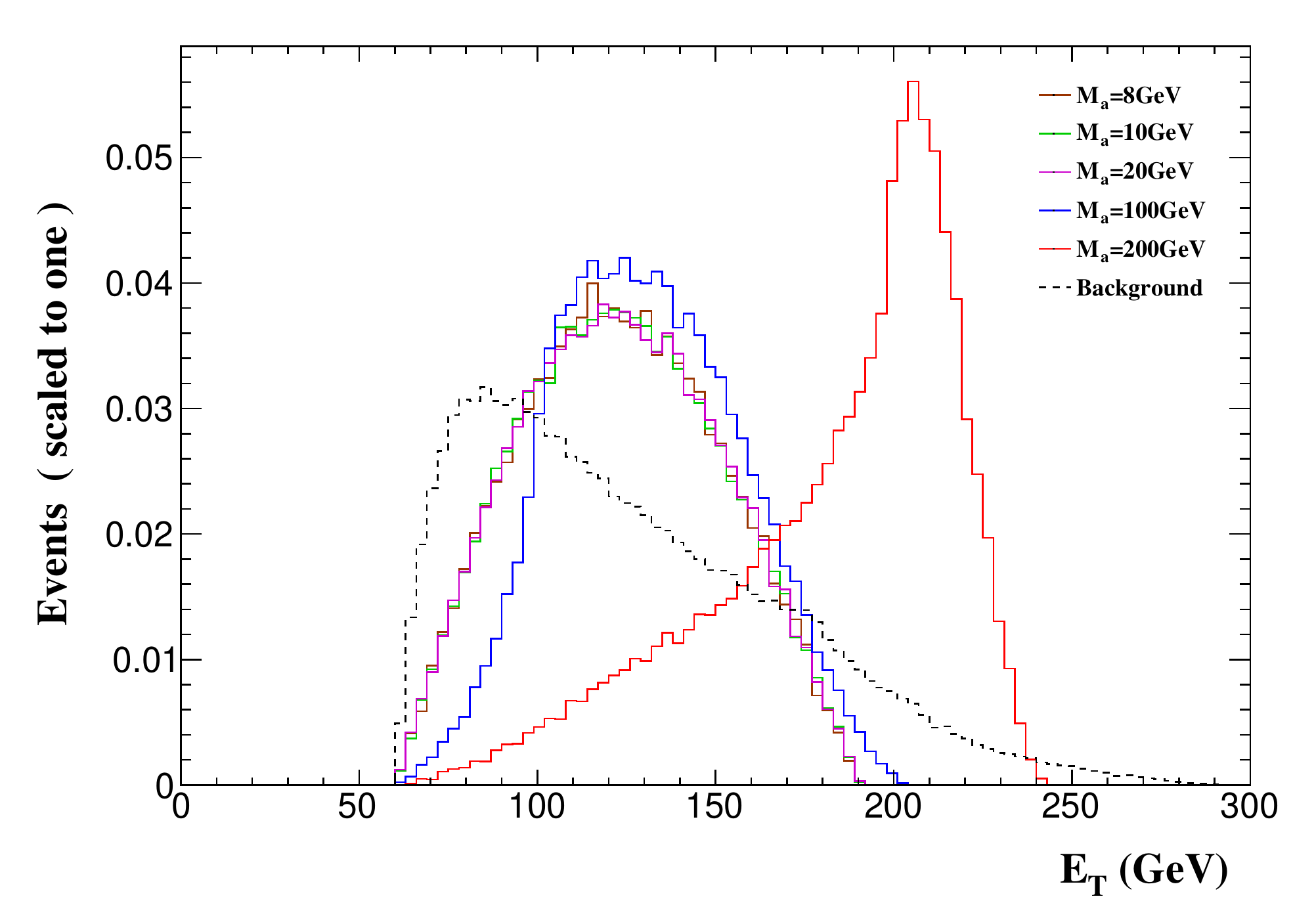}
	\end{subfigure}
	\caption{ Normalized distributions of $M(\gamma~\gamma)$, $\theta(\gamma~\gamma)$, $\slashed{E}_{T}$, $E_{T}$ for the signal of selected ALP-mass  benchmark points and SM background at the $\sqrt{s}$ = 380 GeV CLIC with designed luminosity. }
	\label{CLIC_fig380}
\end{figure*}

\newpage
\vspace{5mm}

\begin{figure*}[!h]
	\centering
	\begin{subfigure}{0.37\linewidth}
		\includegraphics[width=\linewidth]{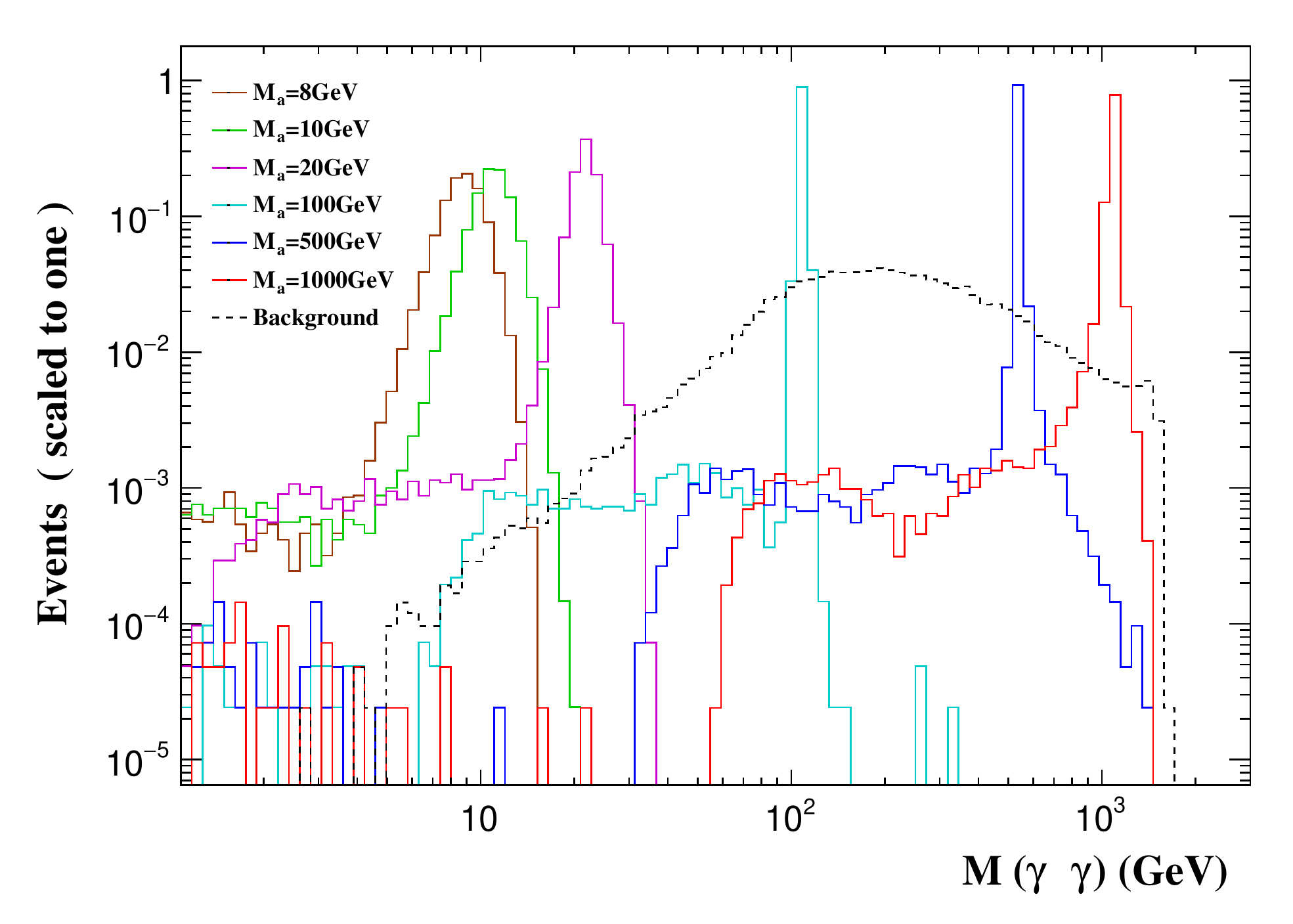}
	\end{subfigure}
	\begin{subfigure}{0.355\linewidth}
		\includegraphics[width=\linewidth]{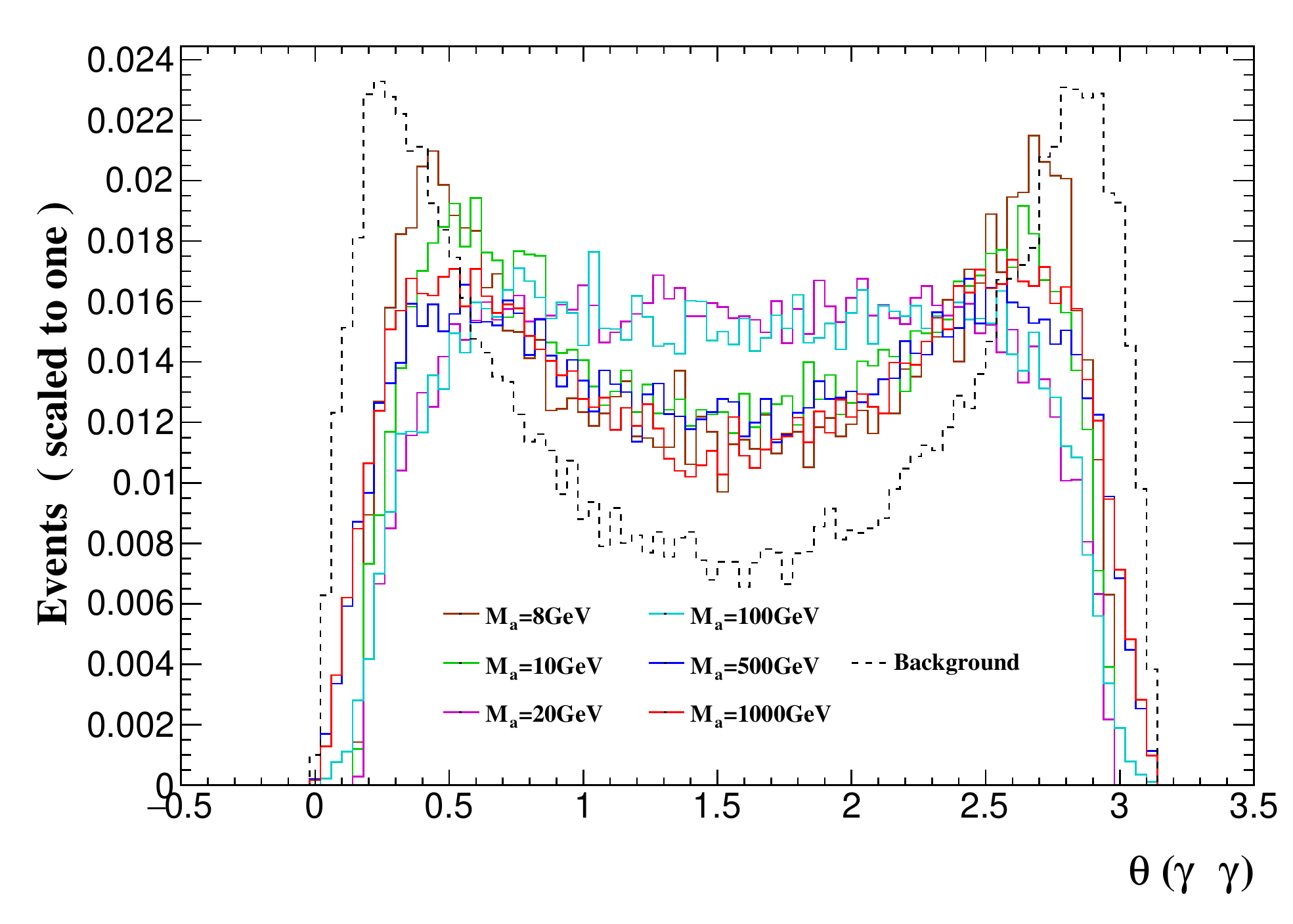}
	\end{subfigure}
	\begin{subfigure}{0.355\linewidth}
		\includegraphics[width=\linewidth]{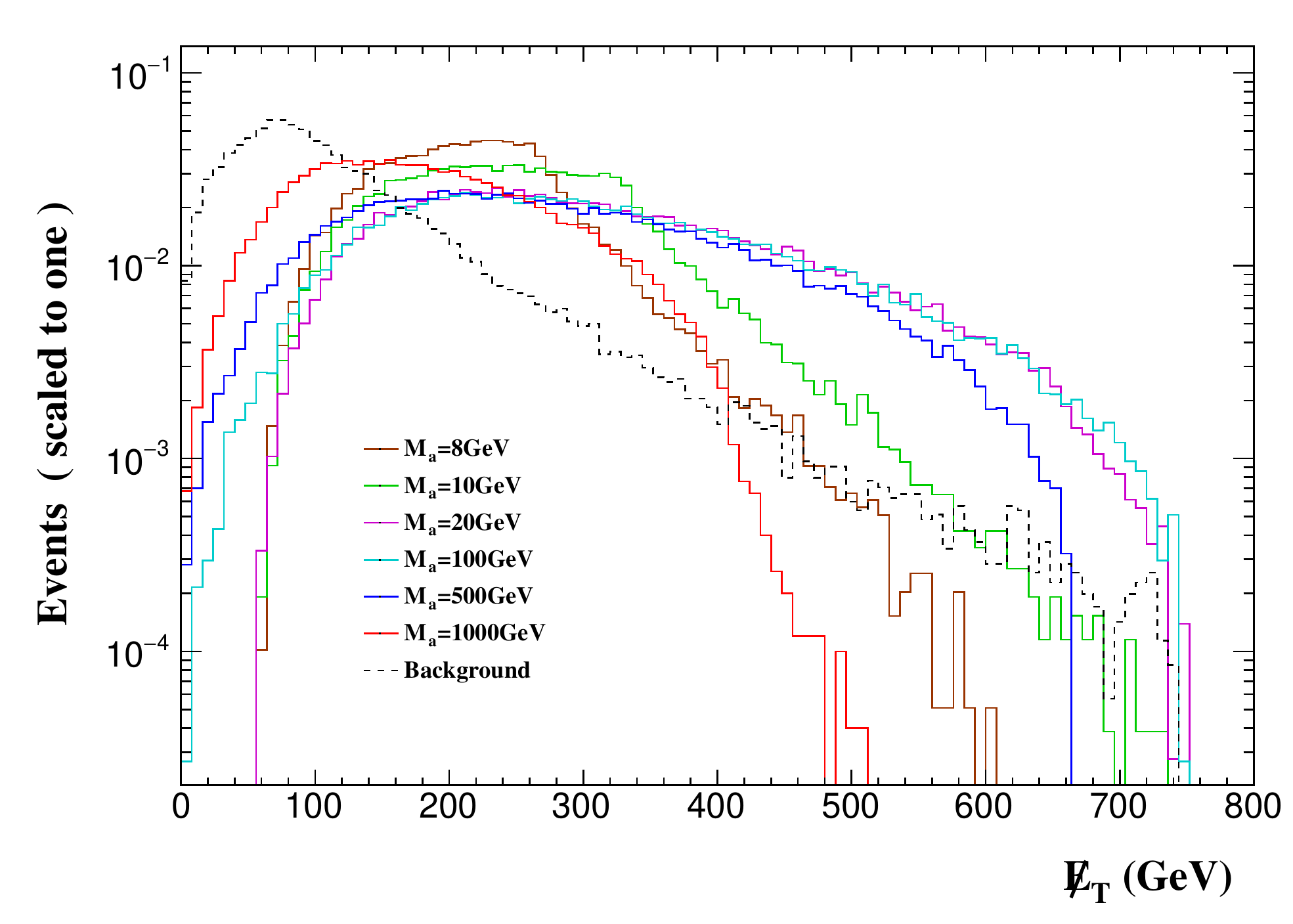}
	\end{subfigure}
	\begin{subfigure}{0.355\linewidth}
		\includegraphics[width=\linewidth]{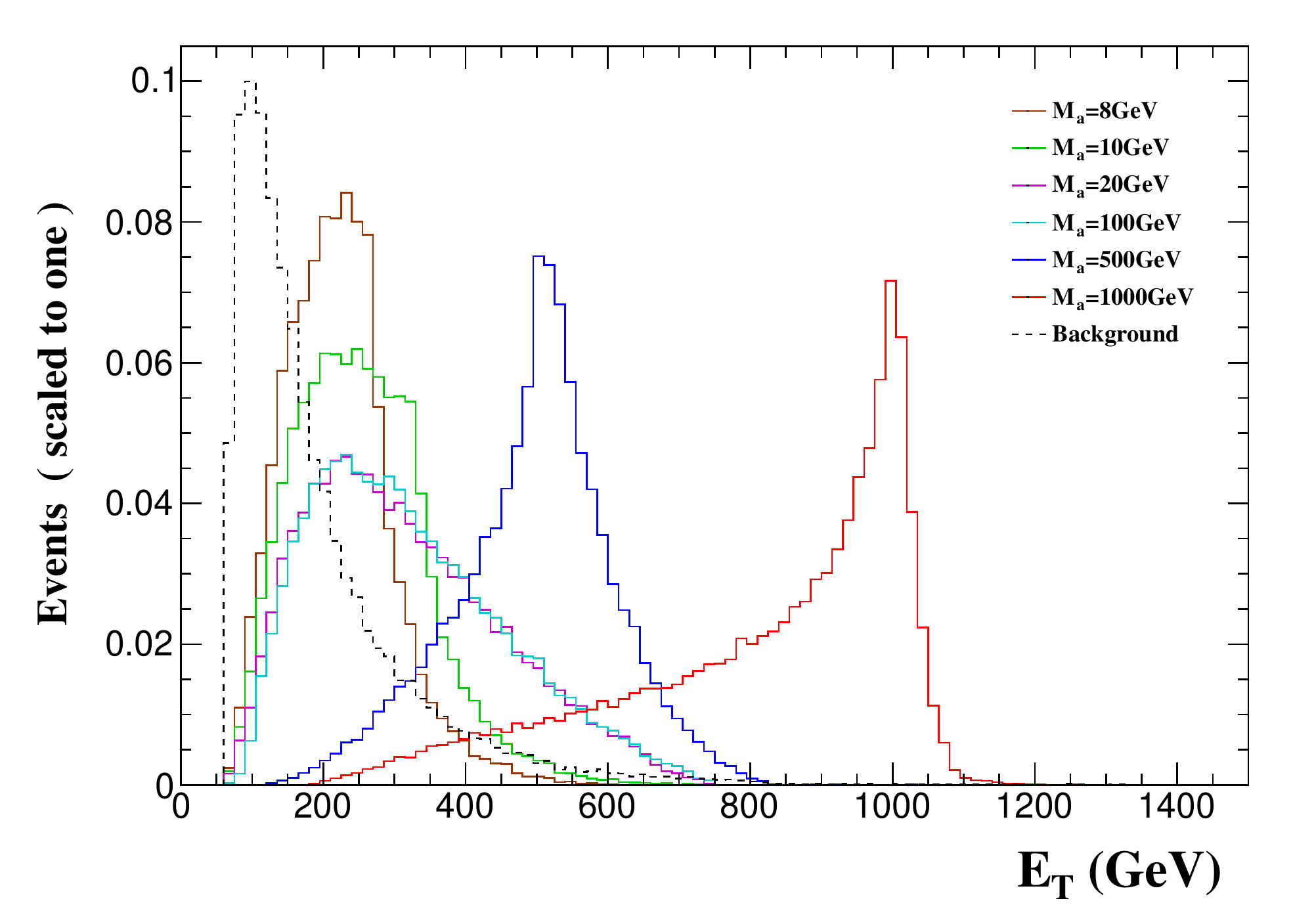}
	\end{subfigure}
	\caption{Same as Fig. \ref{CLIC_fig380} but for $\sqrt{s}$ = 1500 GeV CLIC with designed luminosity.}
	\label{CLIC_fig1500}
\end{figure*}

 \vspace{10mm}

\begin{figure*}[!h]
	\centering
	\begin{subfigure}{0.37\linewidth}
		\includegraphics[width=\linewidth]{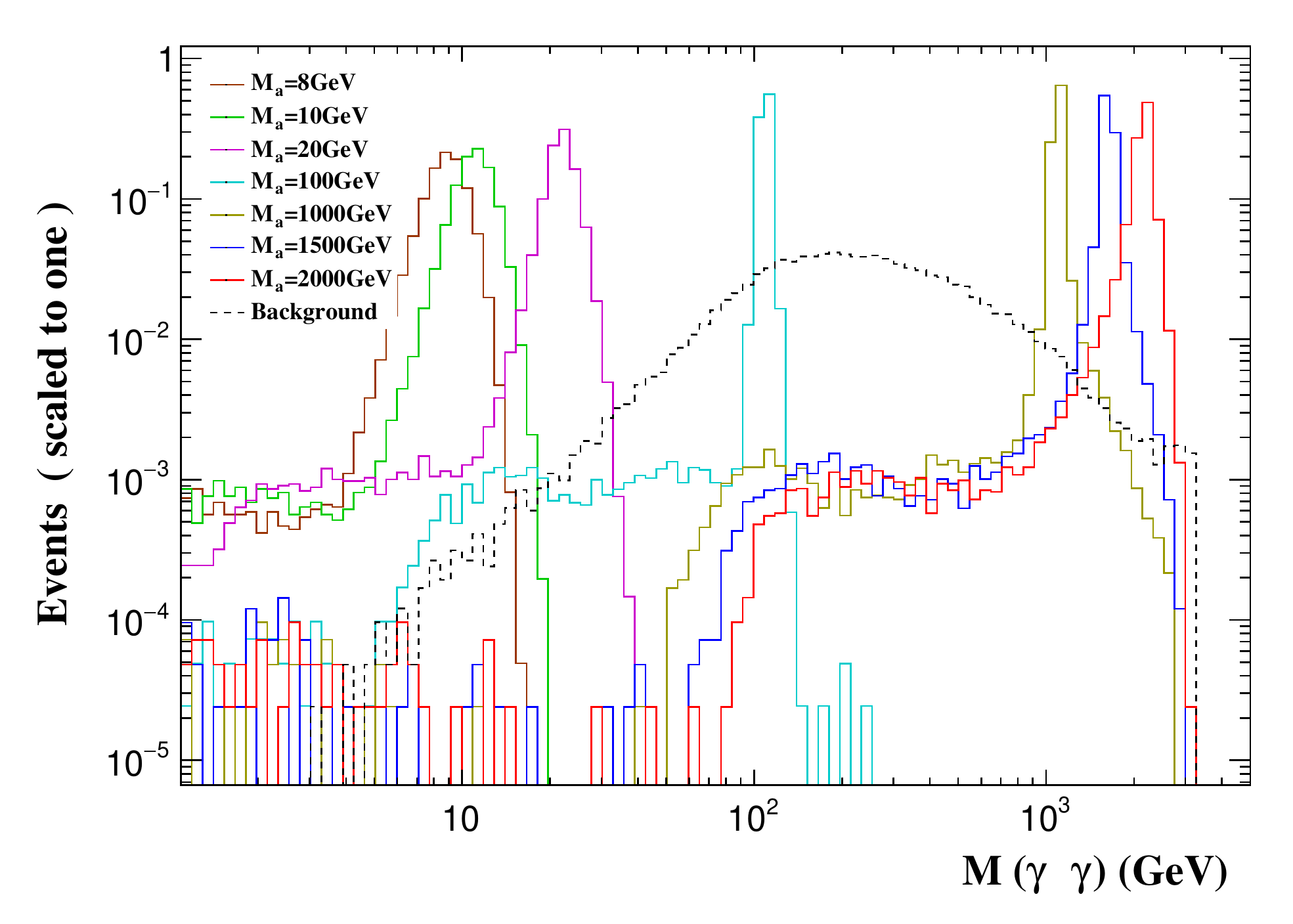}
	\end{subfigure}
	\begin{subfigure}{0.355\linewidth}
		\includegraphics[width=\linewidth]{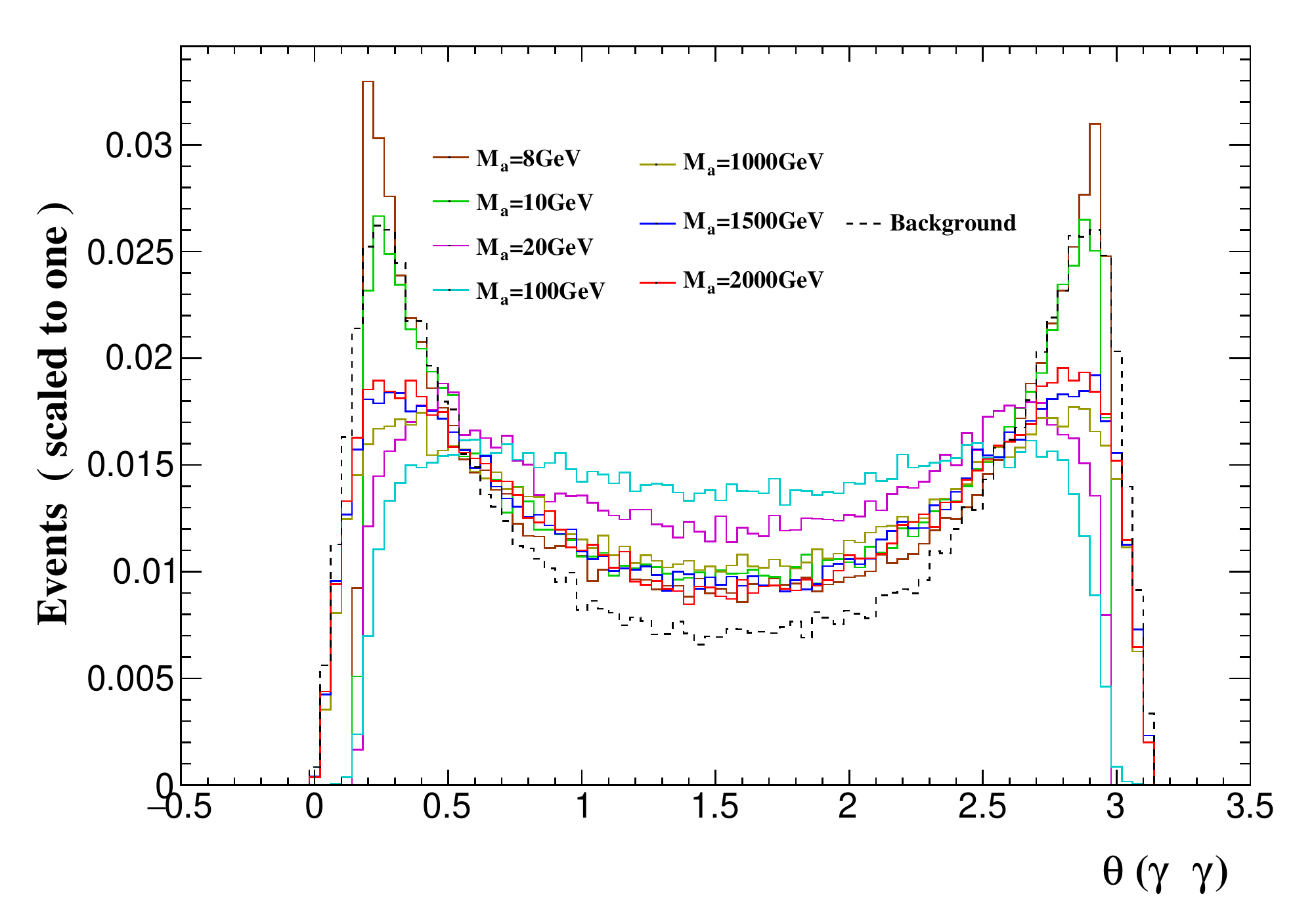}
	\end{subfigure}
	\begin{subfigure}{0.355\linewidth}
		\includegraphics[width=\linewidth]{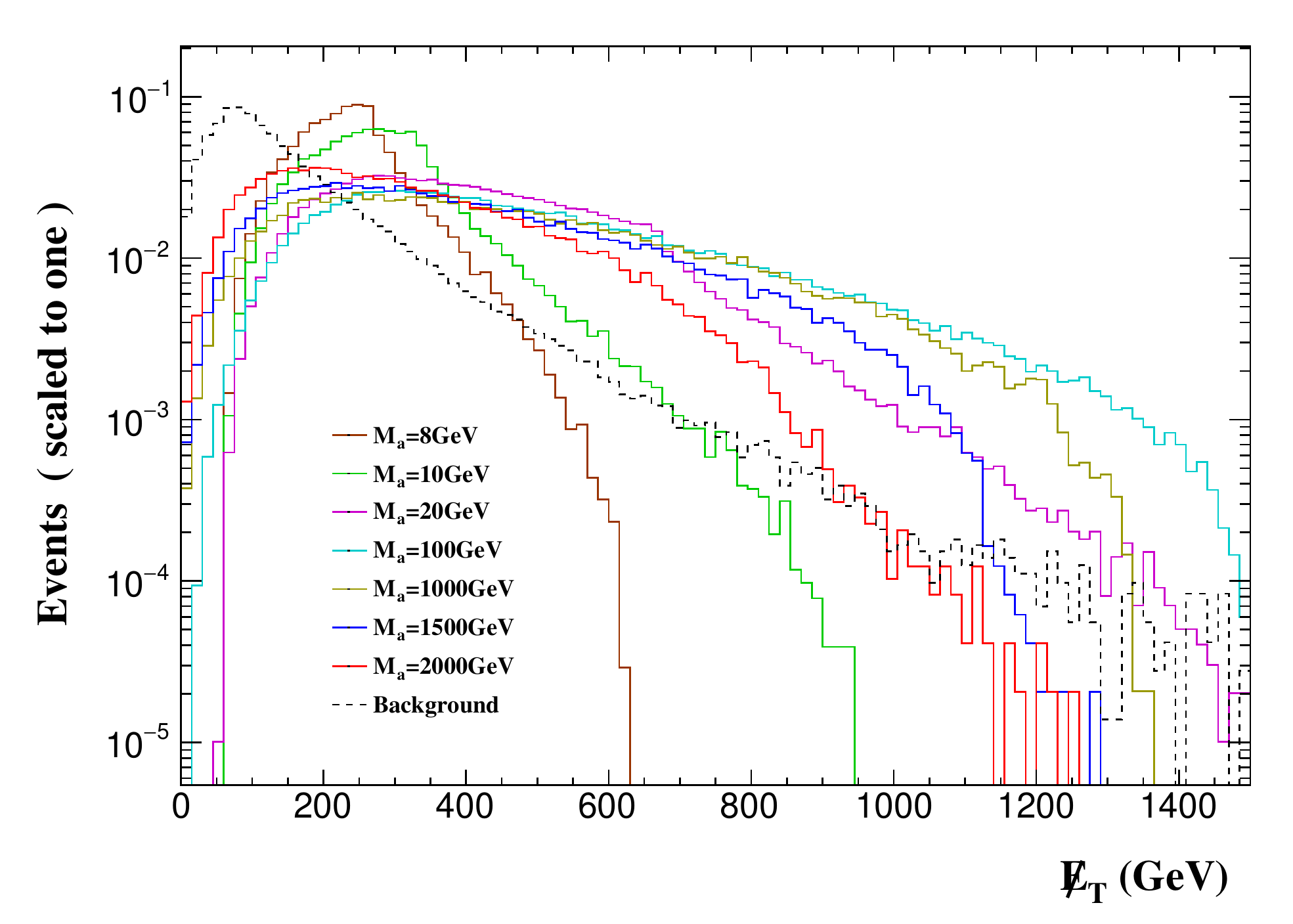}
	\end{subfigure}
	\begin{subfigure}{0.355\linewidth}
		\includegraphics[width=\linewidth]{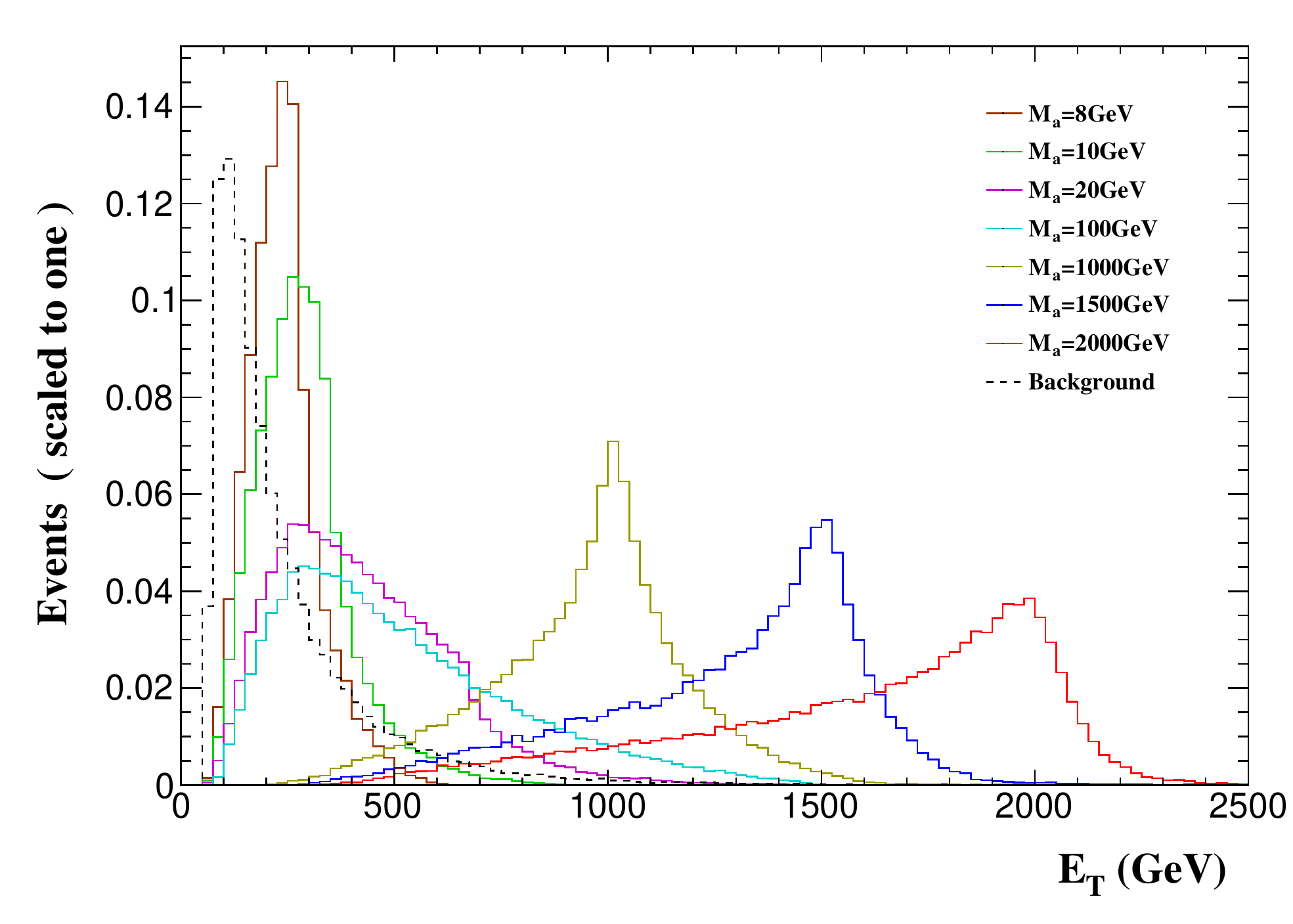}
	\end{subfigure}
	\caption{Same as Fig. \ref{CLIC_fig380} but for  $\sqrt{s}=$ 3000 GeV CLIC with designed luminosity.}
	\label{CLIC_fig3000}
\end{figure*}

\newpage
\vspace{5mm}
\begin{figure*}[h]
	\begin{center}
		\includegraphics [scale=0.5] {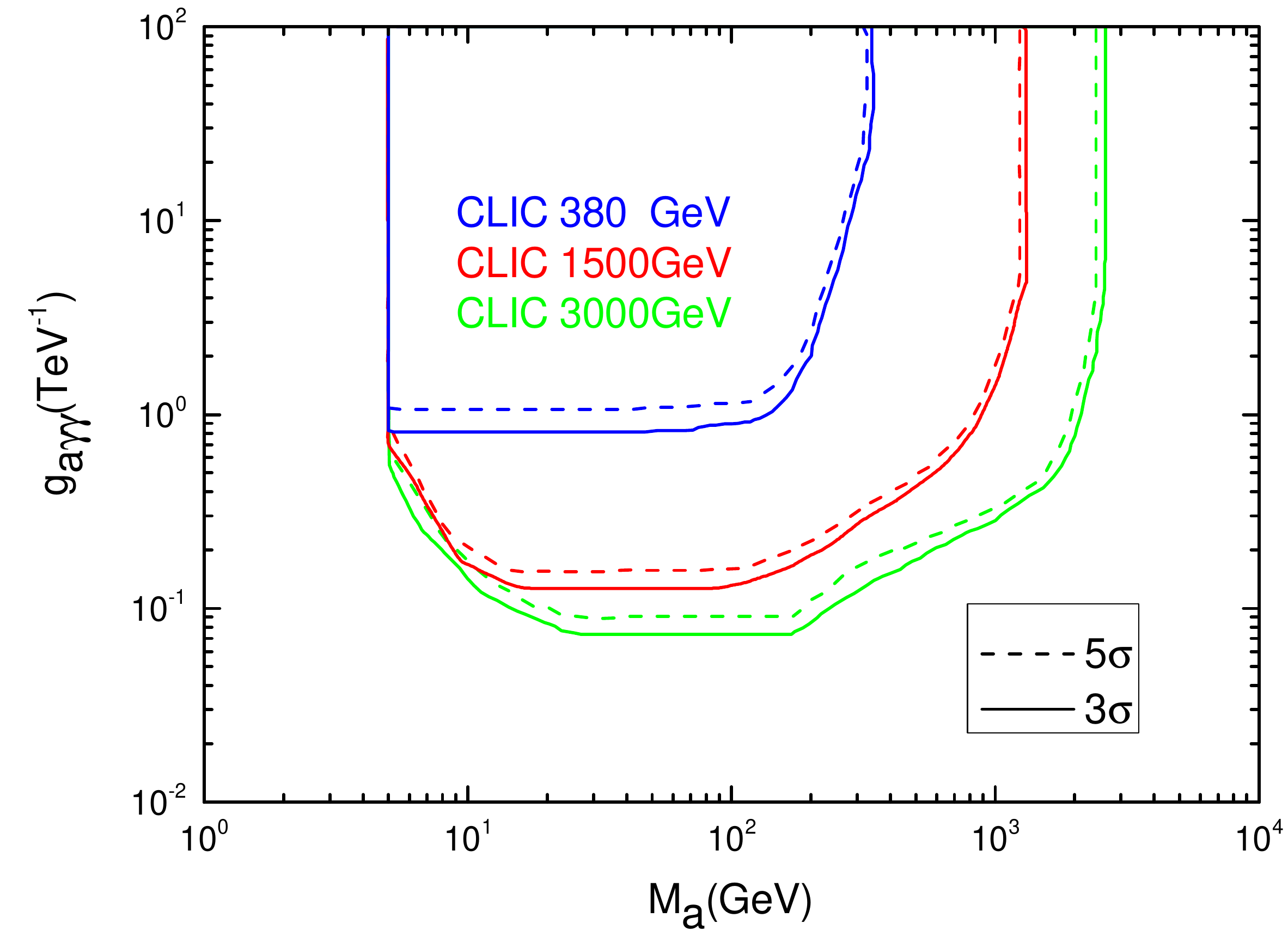}
		\caption{The 3$\sigma$ and 5$\sigma$ curves in the $M_{a}-g_{a\gamma\gamma}$ plane from the W$^{+}$W$^{-}$ fusion process e$^{+}$e$^{-}$ $\rightarrow$ $\overline{\nu}$$_{e}$$\nu$$_{e}a$ $(\rightarrow \gamma \gamma)$ at the 380, 1500 and 3000 GeV CLIC with the designed luminosities. }
		\label{3-5sigmavs}
	\end{center}
\end{figure*}

\vspace{10mm}

\begin{figure*}[!h]
	\begin{center}
		\includegraphics [scale=0.5] {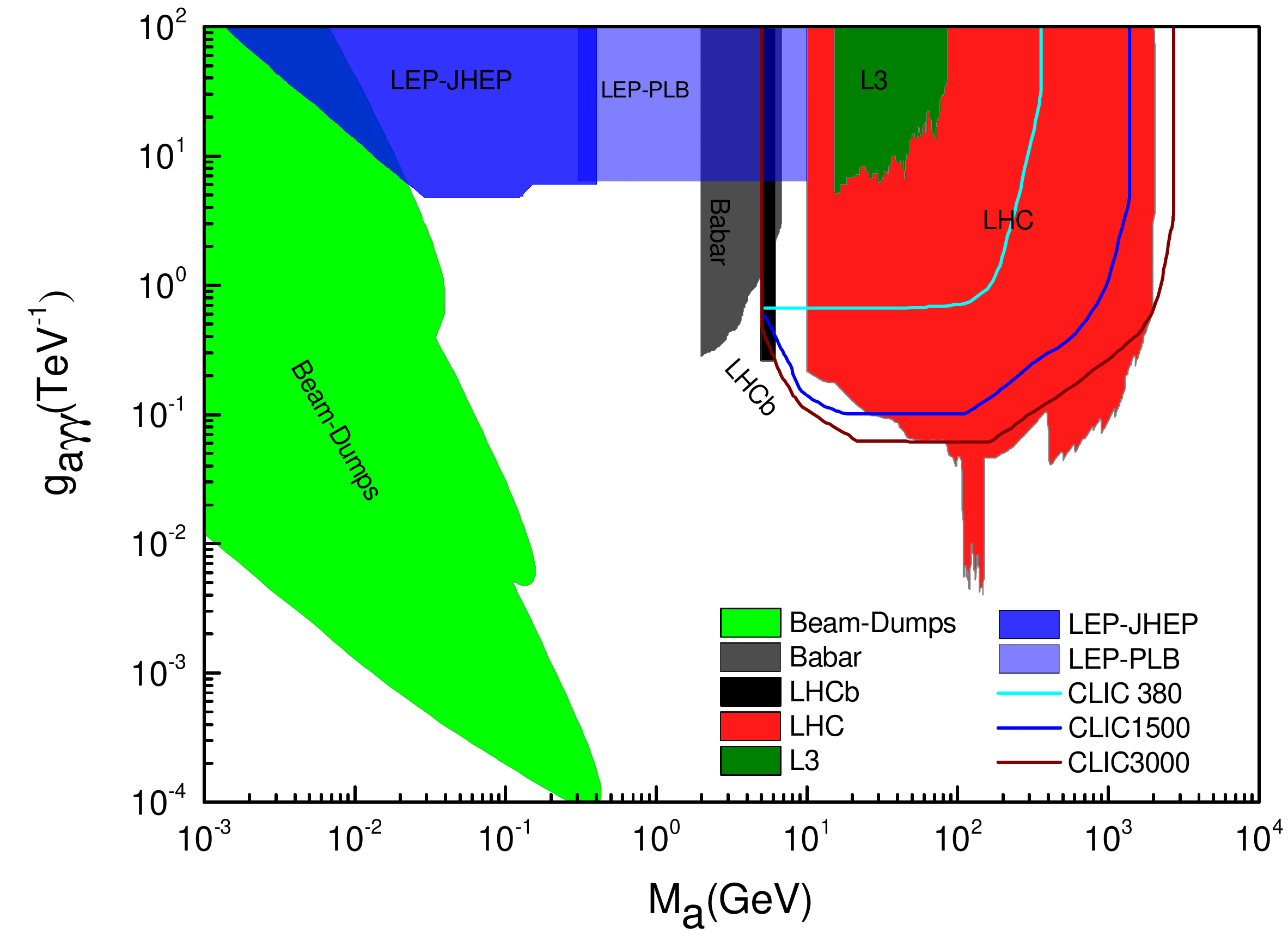}
		\caption{The 2$\sigma$ exclusion limits on the ALP couplings $g_{a\gamma\gamma}$ as a function of $M_{a}$ from the W$^{+}$W$^{-}$ fusion process$~$e$^{+}$e$^{-}$ $\rightarrow$ $\overline{\nu}$$_{e}$$\nu$$_{e}a$~$(\rightarrow \gamma \gamma)$ $~$and other current exclusion regions.}
		\label{2sigmavs}
	\end{center}
\end{figure*}

\end{appendices}
\end{onecolumn}
\end{document}